\documentclass[12pt]{article}

\usepackage[utf8]{inputenc}
\usepackage[T1]{fontenc}
\usepackage[margin=2.6cm]{geometry}
\usepackage{microtype}
\usepackage{graphicx,caption,subcaption}
\usepackage{amsmath,amsthm,amsfonts,amssymb}
\usepackage{enumitem}
\usepackage{fouriernc}
\usepackage{xcolor}

\usepackage{xurl}

\usepackage[hidelinks]{hyperref}
\hypersetup{
    colorlinks,
    citecolor = {red!90!black},
    urlcolor = {blue!80!black}}

\usepackage{orcidlink}
\usepackage{natbib}

\title{Political Stress Index of Poland}

\author{
Tomasz Stachowiak\orcidlink{0000-0001-9851-9131}$^1$
\; and \;
Zbigniew Pasek\orcidlink{0000-0003-2580-4366}$^2$
}

\footnotetext[1]{Independent Scholar, Krak\'ow, Poland;\; \texttt{tomasz@monodromy.group}}
\footnotetext[2]{AGH University of Krakow, Krak\'ow, Poland;\;
\texttt{pasek@agh.edu.pl}}

\date{}

\begin{document}
    
\maketitle

\begin{abstract}
\noindent We apply the political stress index as introduced by \cite{Goldstone} and implemented by \cite{Turchin:2013}, to the case study of Poland. The approach quantifies political and social unrest as a single quantity based on a multitude of economic and demographic variables. The present-day data allow us to directly apply index without the need of simulating the elite component, as was done previously. Neither model version shows appreciable unrest levels for the present, while the simulated model applied to partial historical data yields the index in remarkable agreement with the fall of communism in Poland.
We next analyze the model's sensitive dependence on its parameters (the hallmark of chaos), which limits its utility and application to other countries. The original equations cannot, by construction, describe the elite fraction for longer time-periods; and we  propose a modification to remedy this problem. The model still holds some predictive power, but we argue that some components should be reinterpreted if one wants to keep its dynamical equations.
\\

\end{abstract}

\newpage

\section{Introduction}
\begin{quote}
\textit{In about 6,500 years, if current growth continues,
the descendants of the present world population would form
a solid sphere of live bodies expanding with a radial velocity that,
neglecting relativity, would equal the velocity of light.}
\hspace*{\fill}\citep{Coale}
\end{quote}

\noindent While Asimov's psychohistory remains an unattainable dream, the present day large-scale, global access to vast quantities of data combined with the (so-far) exponentially increasing computational capabilities, and ever increasing sophistication of the accompanying algorithms offer new hopes, or temptations, of forecasting human behaviour. Even if individuals remain safely unpredictable, the decisions of large social groups could -- by analogy with thermodynamics -- prove amenable to mathematical analysis.

The growing scientific specialisation together with all the above reasons, yields a variety of attempts of different backgrounds: from standard game theory \citep{Neumann}, causal graphs \citep{Pearl}, through (co)evolutionary dynamics \citep{Dieckmann, Doebeli} to agent simulations \citep{Lustick, Smaldino} to superforcasting \citep{Tetlock}. The critical voices can be found both urging to simplify \citep{Stupid} and to expand \citep{Elephant}. In order to contribute a solid result,
this article will study one such model, quantifying social unrest or political stress specifically with a view to predicting crises.
The main goal will be the application of the theory and model proposed by \cite{Turchin:2013} to the case study of Poland.
Since the model correctly predicted the (then) coming unrest in the United States \citep{Turchin:2020}, and seems to be the only directly applicable framework, it is ideal to ascertain Poland's situation. Especially since it is already manifesting various signs of social unrest and decline of democracy.
It would also provide a second detailed analysis of Goldstone's and Turchin's ideas, with the aim of further improving the general theory.

The Turchin's approach that we will analyze comes from a long research tradition, going back at least to Max Weber's thought. The basic terms of the theory, such as mass mobilisation, can be found in publications from the second half of the 20th century: 
\cite{Skocpol}, \cite{Goldstone}, and \cite{Collins}. Its crucial element is the intra-elite competition that is an overarching theme in Turchin's analysis of historical crises \citep{Turchin:2009}.

Our results are threefold. First, regarding the current situation in Poland, both the original model and the one based on measured elite data do not indicate significant levels of unrest. Second, the discrepancy between the data and model led us to a deeper mathematical analysis, which shows that the equations are not well-posed, and they either have to be augmented or the interpretation of the variables has to change; we propose one such reinterpretation. Third, in an attempt to further test and calibrate the model, we applied it to the incomplete historical data (since 1950), and the alignment with the fall of communism in Poland, despite many caveats, suggests that the model deserves further attention and refinement.

\section{Political Stress Index}
\label{main_sec}

\begin{quote}
\textit{To those seeking a cause, yet unwilling to accept any hypothesis of reason, be it providential or devilish in form, only the rational surrogate of demonology remains -- statistics.}\\
\hspace*{\fill} Stanis\l{}aw Lem, \textit{G\l{}os Pana}
\end{quote}

\noindent We shall start with stating how the main indicator of unrest is constructed going top-down, so that it will immediately be clear where direct data can be used, and where some theoretical intervention is required -- be it modelling, extrapolation or an educated guess.

The components of political stress, as identified by Goldstone, are: the Mass Mobilisation Potential (MMP), Elite Mobilisation Potential (EMP) and State Fiscal Distress (SFD) \citep{Goldstone}. That they are factors contributing to unrest, revolutions, or civil wars is intuitively clear, but in order to capture the information they carry, they have to be combined into a single Political Stress Index (PSI), while each component has to be either measured or further decomposed into measurable quantities.

Goldstone proposes that the former step be achieved through the simple formula
\begin{equation}
    \Psi = MMP \cdot EMP \cdot SFD,
    \label{main_psi}
\end{equation}
which gives equal weights to the three variables. This is not the only choice, but it does have the desirable trait of reflecting changes in orders of magnitude of each variable while the others are held equal. Though simple, such an equation is a good starting place for modelling a phenomenon for which we so far have no fundamental theory.

Each component in \eqref{main_psi}, has likewise many possibilities. We will adopt the approach that was proposed by \cite{Turchin:2013}, for it relies on simple principles and accessible data (at least for the recent past). As before, each variable will have a multiplicative form, built directly from macroeconomial indicators. Namely:
\begin{equation}
\begin{aligned}
MMP &= \frac{N_{\text{urb}} \cdot A_{20-29}}{w},\\
EMP &= \frac{e}{\epsilon},\\
SFD &= y \cdot D,
\end{aligned}
\end{equation}
where the basic quantities are:
\begin{itemize}
    \item $w$: wages divided by the GDP per capita,
    \item $N_{\text{urb}}$: fraction of the population living in cities (urbanisation),
    \item $A_{20-29}$: fraction of the population aged 20-29 (youth bulges),
    \item $e$: fraction of elites in the society
    \item $\epsilon$: elite's wages divided by the GDP per capita,
    \item $y$: government debt divided by the GDP,
    \item $D$: public distrust in the government institutions.
\end{itemize}

As we will see, there are two modes of proceeding with the ingredients. Some of them, like population, are readily available for many years into the past; while for others, like the distrust or elite numbers, we only have a few years of recent polls, not to mention definitional difficulties. Turchin remedies this problem by constructing dynamical models to supplant the data, and verifying their correspondence with the past (The American Civil War for example).

Although we will be able to clearly follow all the variables for the past 30 years in Poland's case, the distant-past data will have to be accepted provisionally, due to the nature of the communist regime since the 1950's up to 1991. Most of the data were gathered yearly, and the straight lines in the graphs are added just for readability. Unless otherwise stated, such linear interpolation is also used for missing data-points when incompatible data sets have to be combined.

More importantly, by trying to model the situation with Turchin's tools, we have discovered difficulties that must be studied more closely. Partly, they follow from a questionable formulation of the elite model; and partly they show that a model that works for the US, can fail in Poland (or other European countries perhaps) -- the American mechanism that fuels mass-elite dynamics is not universal, and the obtained PSI not as well-founded.

\subsection{MMP}

When it comes to the contemporary data, the Mass Potential is the most straightforward to obtain. The average wages and GDP are regularly published, and corrected by GUS (the governmental statistics agency, currently called Statistics Poland in English), whose data we use. The exact sources of all data are listed in Appendix~\ref{ListApp}.

Problems begin with population numbers, due to changing methodology (e.g. in 2010) as indicated in the macroeconomic tables of GUS. The potential errors propagate to several variables: GDP per capita, urbanisation, and anything else that requires rescaling by the total population. However, we note, following \cite{Turchin:2013}, that the value of $\Psi$ and its three main ingredients are relevant to us only relatively, and that significant signals are reflected through order-of-magnitude changes. The aforementioned gap in the population data is roughly 300 thousand in 38 million, or $0.8\%$.

\begin{figure}[h]
    \centering
    \includegraphics[width=.95\textwidth]{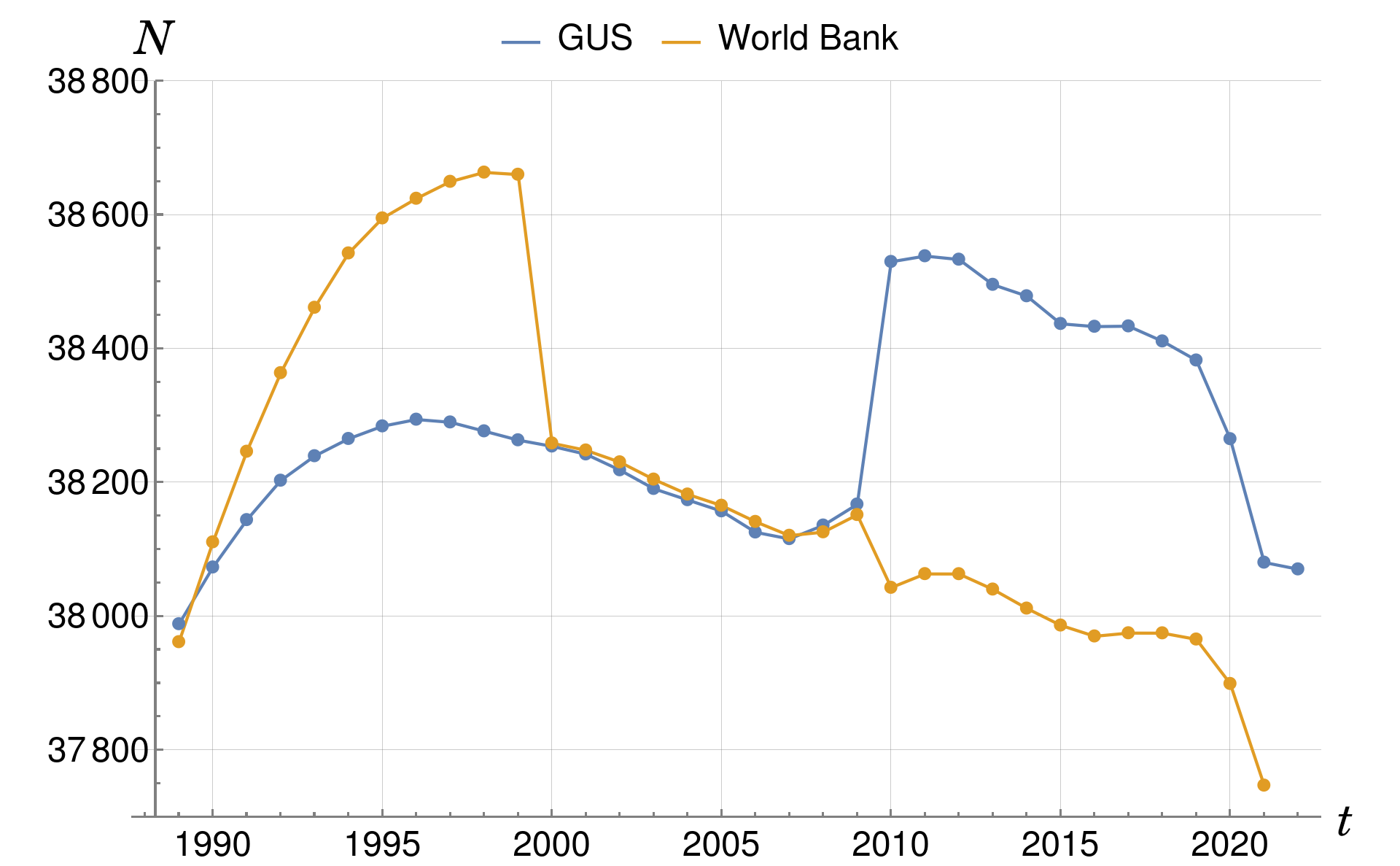}
    \caption{Total population in thousands.}
    \label{fig:population}
\end{figure}

Since this is negligible across the overall dynamics, we choose to use the raw GUS data, keeping the caveat in mind, as it applies again and again. Fortunately, as mentioned in the introduction, we are concerned with a tentative qualitative theory, and will soon find a more troubling source of large errors that cannot be neglected.

\begin{figure}[!htp]
    \centering
    \includegraphics[width=.89\textwidth]{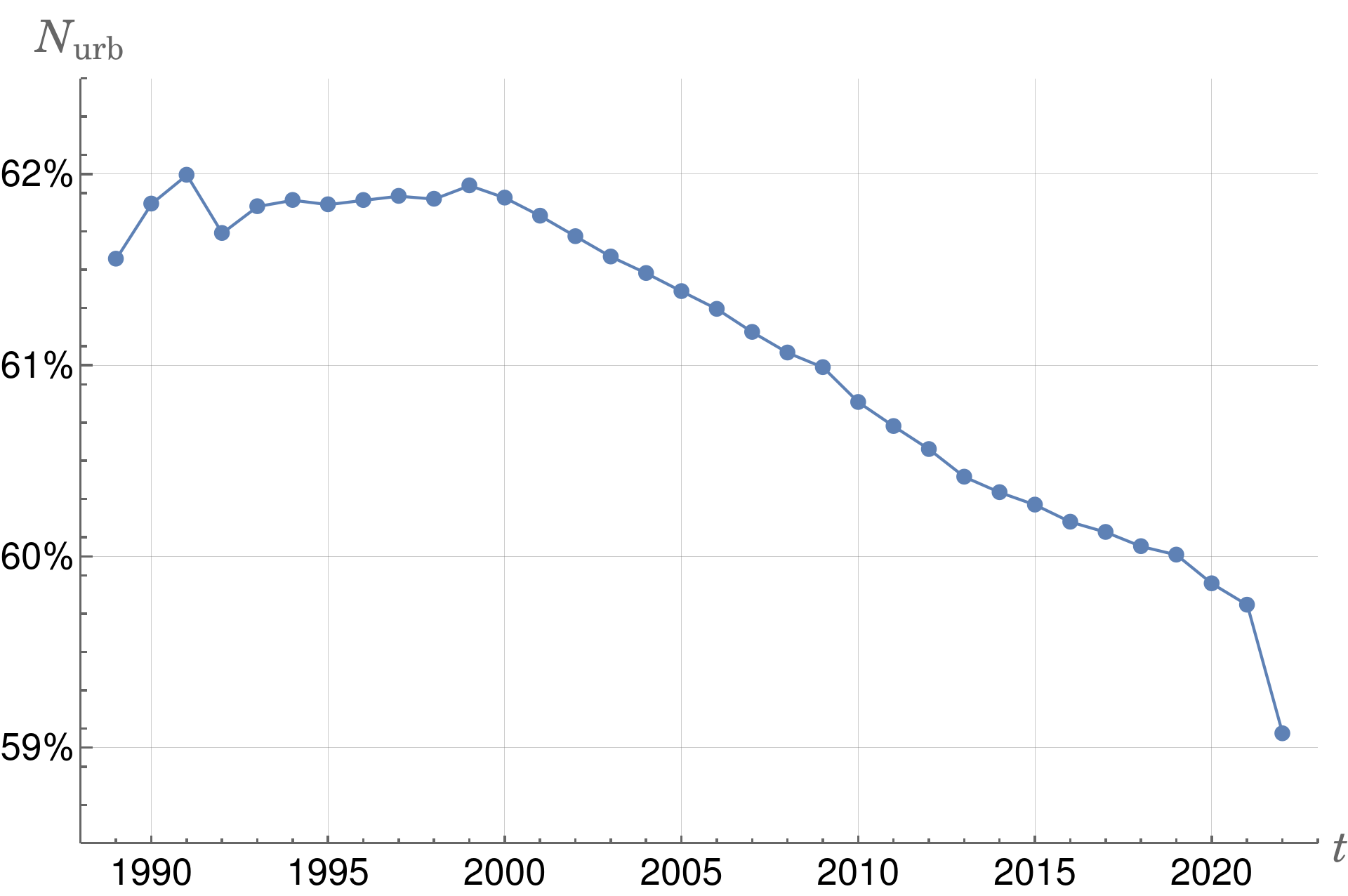}
    \caption{The percentage of the population living in cities (urbanisation).}
    \label{fig:urbanization}
    \includegraphics[width=0.89\textwidth]{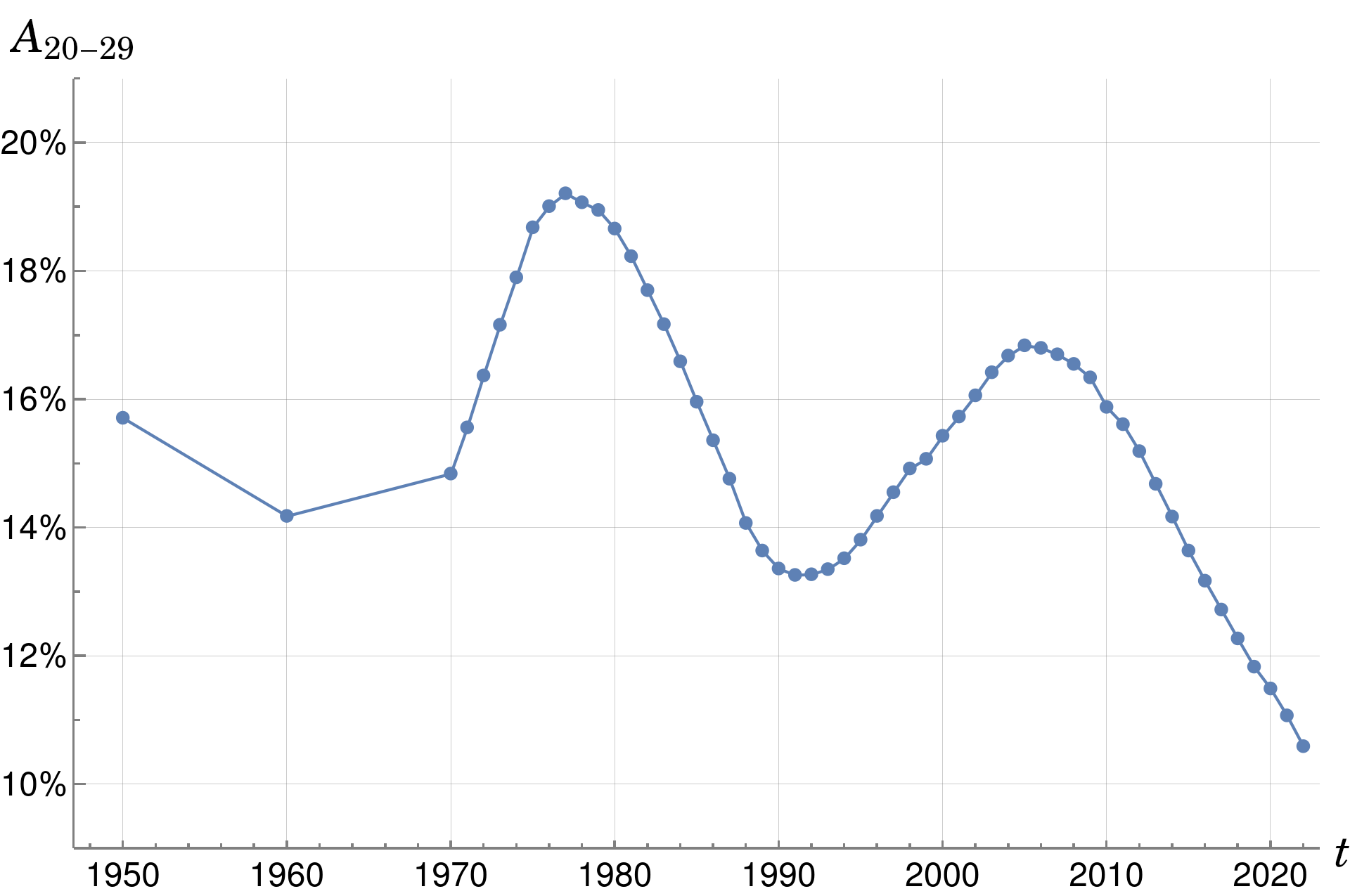}
    \caption{Percentage of people in the 20--29 age group.}
    \label{fig:bulges}
\end{figure}

The urbanisation data are provided by GUS separately, and only in one version, although necessarily the same provision applies regarding the total population. Figure \ref{fig:urbanization} shows the most recent data.

The youth bulges numbers were taken from the Age Pyramid of GUS. Here, the data also include modelling of the population by GUS, to obtain the cohort numbers between the years when National Censuses were conducted, and up to the year 2060. For the details, the reader is directed to the official GUS methodology. Owing to the accessibility of data, we present the extended period of 1950--2023, which allows for clear identification of the ``bulges''.

At this stage of data gathering, it became apparent that almost all of the other indicators could also be extended into the past, and perhaps the whole communist period could be analyzed. For now, we will proceed with the most recent, and reliable, data, and the extended analysis can be found in Section \ref{PRL}.

\begin{figure}[!ht]
    \centering
    \includegraphics[width=.85\textwidth]{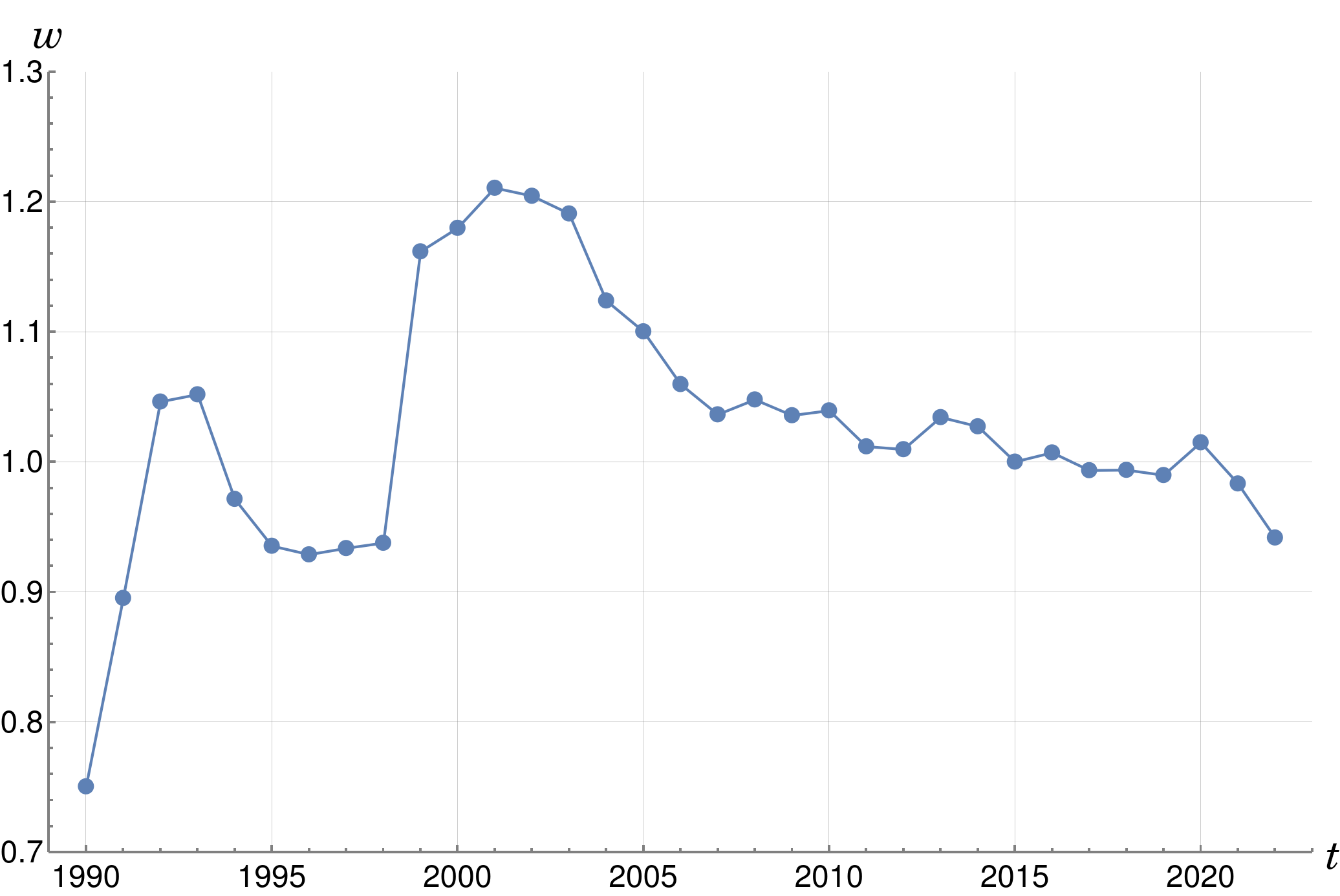}
    \caption{The wages relative to GDP per capita.}
    \label{fig:wages}
\end{figure}

To obtain the relative wages, we take nominal values of the wages and the GDP per capita -- their ratio is, by definition, unaffected by inflation, and we use nominal values in what follows in a similar manner. The resulting $w$ is shown in Figure~\ref{fig:wages}.

\begin{figure}[!h]
    \centering
    \includegraphics[width=.85\textwidth]{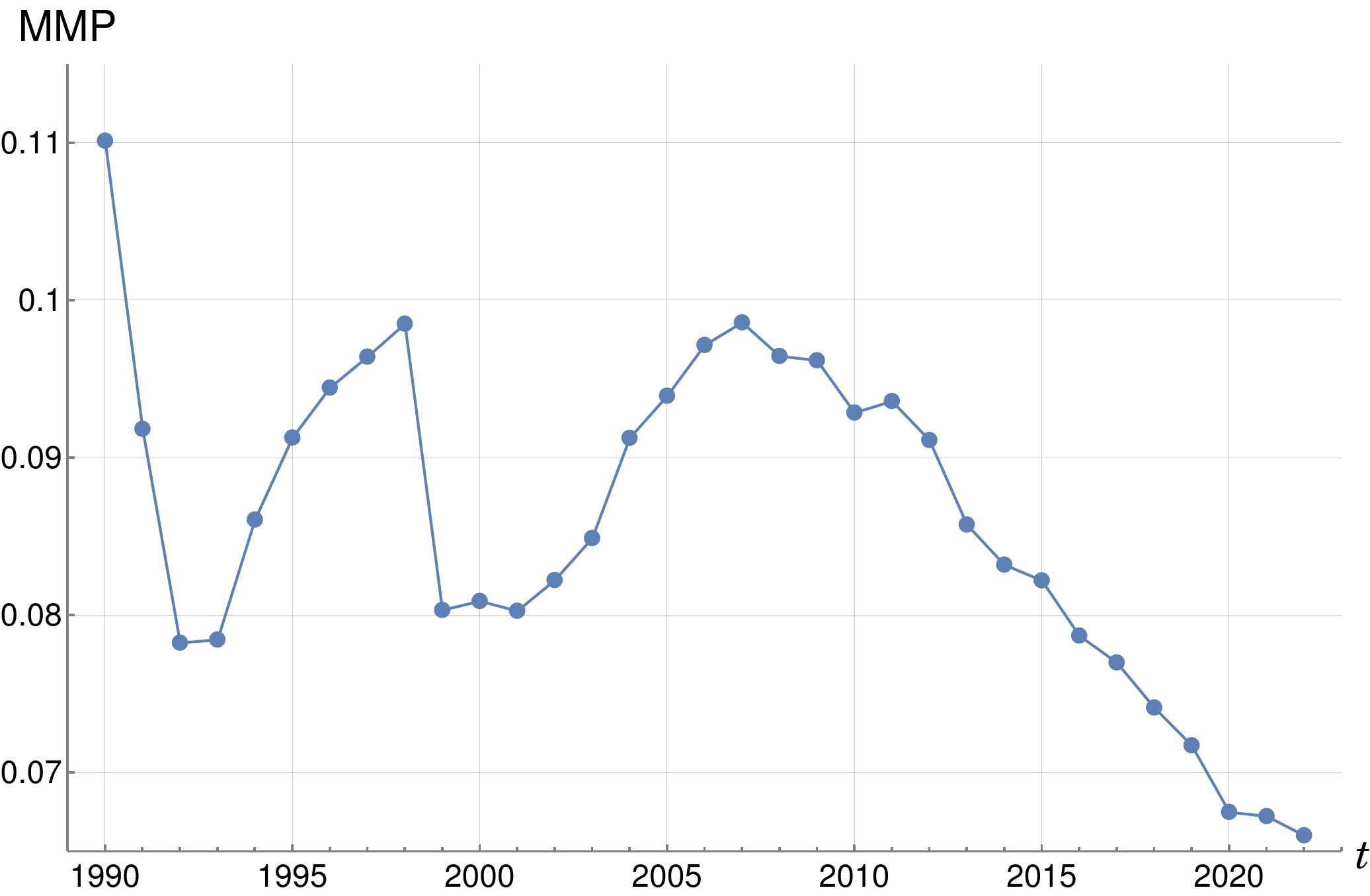}
    \caption{The Mass Mobilisation Potential.}
    \label{fig:mmp}
\end{figure}

The resulting MMP is shown in Figure~\ref{fig:mmp}.
A preliminary comment about its evolution can already be made here, as it is clear that one youth bulge corresponds neatly to the Solidarity movement and the beginning of democracy in Poland. The second such bulge happened around 2006, although no observable revolution followed. One possible explanation is the mass emigration after Poland's accession to the EU (2004) and Schengen (2007) -- the participation of the 20-29 cohort among all emigrants rose from 20\% to over 40\% according to GUS (Appendix~\ref{ListApp}, Population).

\subsection{EMP}
\label{sect:EMP}

This is the most problematic element, due to the many conflicting definitions of ``elite'' as well as the mathematical model in the original paper \citep{Turchin:2013}. Fortunately, we can almost completely circumvent the latter problem: there exist quite detailed data for the Polish labour market published by GUS every two years. This allows for estimation of the elite numbers and their salaries directly, rather than by modeling them based on the population-wide wages. Similar investigations have long been performed by a prominent Polish sociologist, specialising in social stratification: Henryk Doma\'nski \citep{Domanski}, and we follow a similar approach.

Specifically, the GUS labour market data are divided into occupational groups, and we separated them into elites and non-elites. The first group included the categories (as defined by GUS annals): ``managers'', which included public authorities and higher officials as well as CEO's; medical doctors, lawyers, professors (academic teachers), financial and law experts. In addition, 10\% of the following categories were added: physics, mathematics and technical specialists; management, marketing and IT specialists; journalists and artists. The particular names correspond to GUS classification in the structure of average gross wages, as found in GUS Yearbooks of Statistics (see Appendix~\ref{ListApp}). We refer to the result as the ``observed elites''.

This introduces some arbitrariness into the procedure, we accept it due to two reasons mainly. First, we discovered \textit{post hoc}, that the resulting numbers agree quite closely with those of \cite{Domanski}, whose exact methodology is independent. And second, it should be repeated that even in the original model, it is not necessary to use an objective definition of the elite, as the drive of mobility is in fact the \emph{perceived} elite. I.e. people (would like to) move to occupations which they consider to be elite.

This does not solve all problems, as it has long been the case that e.g. nurses and firefighters are rated high in Polish prestige polls \cite{CBOS}, even though they don't come to mind as the elites, nor is there a great influx into those professions. A fuller analysis will necessarily have to make distinctions between intellectual, financial or ruling elites (among others), but here again we limit ourselves in accordance with the original model, where the whole elite component was estimated through the dynamics of wages, so that the financial aspect dominates.

\begin{figure}[!ht]
    \centering
    \includegraphics[width=.90\textwidth]{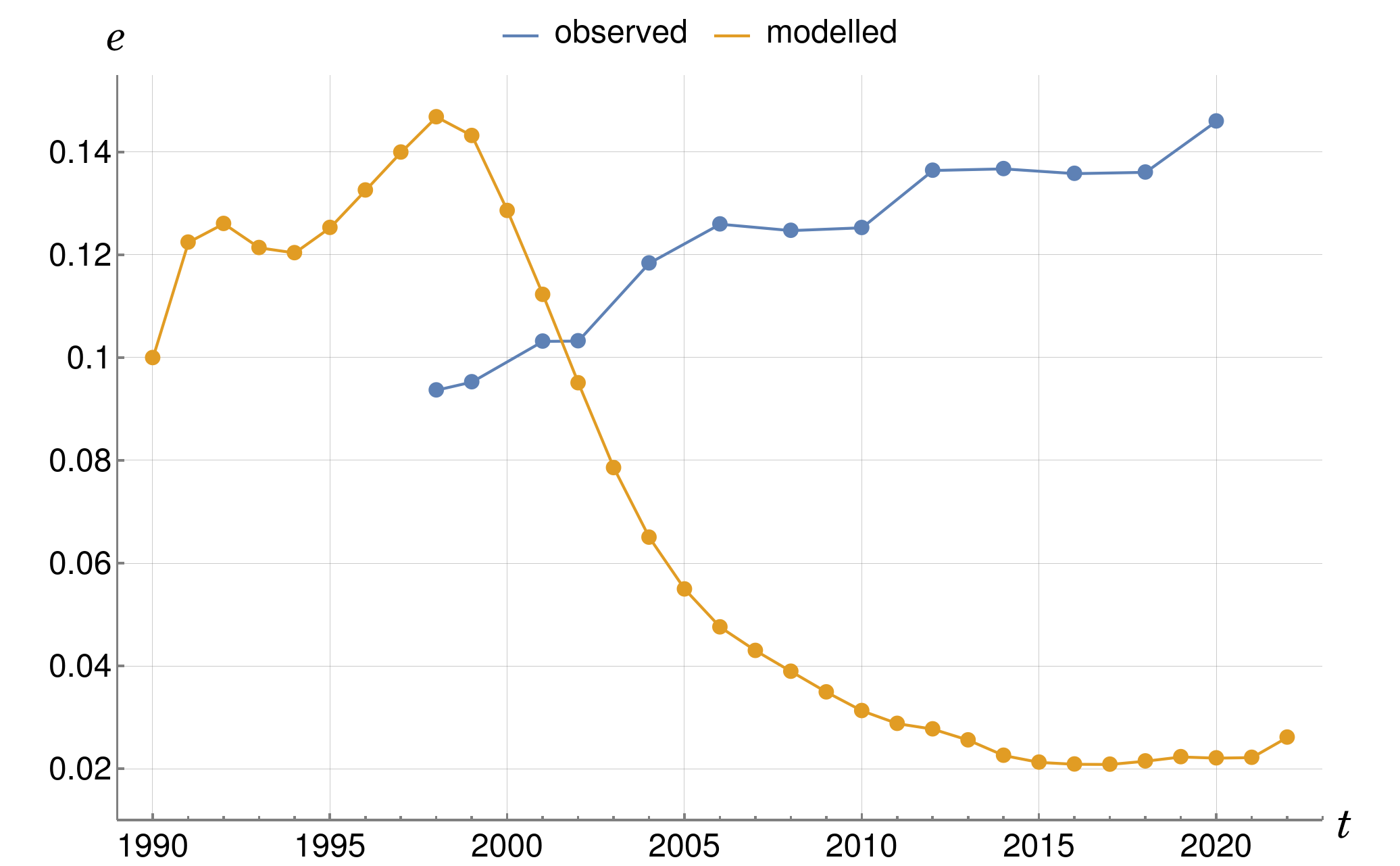}
    \caption{The relative elite numbers: observed (blue) and modelled (orange).}
    \label{fig:elites}
\end{figure}

The observed and modelled elite numbers are presented in Figure~\ref{fig:elites}, and a huge discrepancy is immediately visible. The basic reason behind this is that the original model \cite{Turchin:2013} is purely ``financial'', i.e. it assumes that it is the relative wages that drive the up- and downward mobility from masses to elites and back. Specifically, the absolute populations are governed by the differential equations:
\begin{equation}
    \begin{aligned}
        \dot{N} &= r N\\
        \dot{E} &= rE + \mu N,
    \end{aligned}
\end{equation}
where $E$ are the elites, $N$ is the overall population, $r$ is the per capita rate of natural growth, and $\mu$ a parameter quantifying mobility. This parameter is then assumed to depend on relative wages through
\begin{equation}
    \mu = \mu_0 \frac{w_0-w}{w},
\end{equation}
so that whenever $w$ falls below a set level $w_0$, $\mu$ becomes positive, and people migrate from the masses to the higher earning elites. Finally, since we are interested in the fraction of elites $e=E/N$, the two equations simplify into one:
\begin{equation}
    \dot{e} = \mu_0\frac{w_0-w}{w}.
    \label{eq:ee}
\end{equation}

There is little explanation given in \cite{Turchin:2013} as to the selection of the two parameters $\mu_0$ and $w_0$, even though the whole model is very sensitive to their values. What is worse, the author uses two $\mu_0$ differing by several orders of magnitude: 0.002 and 0.1. In the end we decided to take the values that were used in the final analysis (20th century) in \cite{Turchin:2013}, i.e., $\mu_0=0.1$ and $w_0=1$. We also had to decide on some initial condition for the elite fraction $e$, which we set to be 10\% in 1990. Note, that the whole (orange) graph in Figure~\ref{fig:elites} would only be shifted up or down with the change of this initial condition, and even if $e$ were set to be 10\% in 2020, we cannot escape the unrealistic results: either $e=2\%$ in 2020 (the displayed orange curve), or $e\approx 22\%$ in 1998. Worst of all, if the graphs are set to agree in 1998, the percentage becomes negative in 2020.

The difficulties only get more pronounced when the extended data are included, skipping ahead for a moment to Figure~\ref{fig:full_elites} we learn that Poland had 93\% elites in the 90's. In view of the previous paragraph, trying to shift the graph to reflect the current state, we end up with negative fraction in the past again. Since these are not the only difficulties, we postpone their discussion to Section \ref{critique}, and proceed with the analysis.

The model of elite wages is simpler, based on the idea that not all of GDP is divided between the working masses, and the remainder goes to the elites. This gives the simple formula for their relative wages:
\begin{equation}
    \epsilon = \frac{1-\lambda w}{e},
\end{equation}
where $\lambda$ is the fraction of the total population that is employed, taken to be 0.5 as in the original model.

It must be pointed out that the above setup effectively means counting $e$ twice, because EMP contains the fraction $e/\epsilon\propto e^2$. This will be especially significant in view of our later analysis in Section \ref{critique} -- the whole PSI is dominated by the EMP component, which, in turn, is dominated by the disparity in wages.

\begin{figure}[!ht]
    \centering
    \includegraphics[width=.860\textwidth]{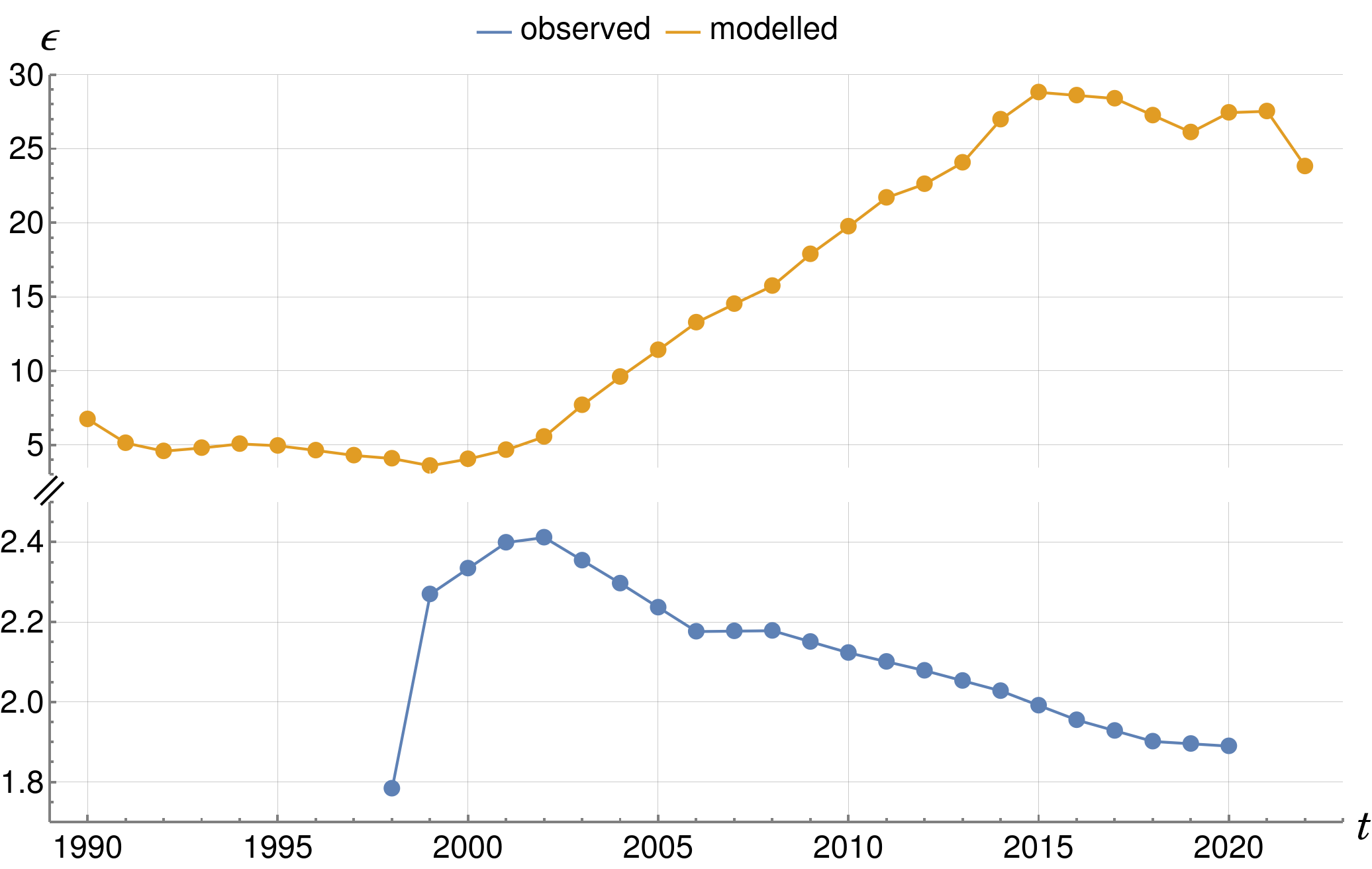}
    \caption{The relative elite wages. The vertical axis is broken, due to very different ranges.}
    \label{fig:epsilon}
\end{figure}

\begin{figure}[!h]
    \centering
    \includegraphics[width=.90\textwidth]{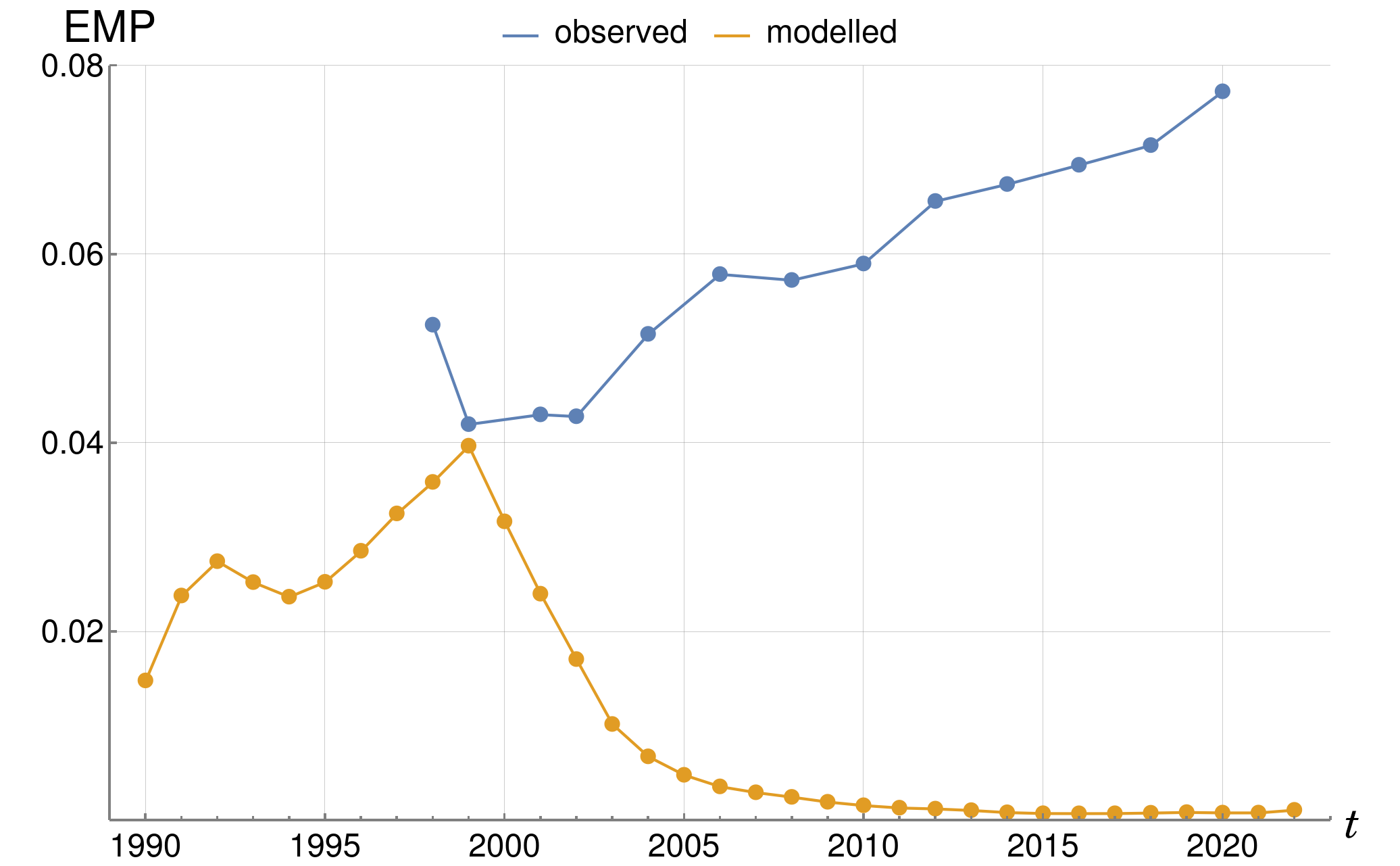}
    \caption{The Elite Mobilisation Potential: observed (blue) and modelled (orange).}
    \label{fig:EMP}
\end{figure}

We, on the other hand, can also use the official GUS data, with explicit salaries for all the aforementioned groups. The observed elite wages are compared with those obtained from the model in Figure~\ref{fig:epsilon}; and, combing the above variables,  we can look at the two versions of the EMP component, which can be found in Figure \ref{fig:EMP}.


The full mathematical analysis of the EMP will be the subject of another section, where we will argue, that it cannot be applied as is. But to fully present the attempted case study, we will use also the original version, i.e., the elite dynamics as obtained from the level of wages and upward mobility. This will consequently provide a new empirical test of Turchin's proposed model.

\subsection{SFD}

\begin{figure}[!htp]
    \centering
    \includegraphics[width=.925\textwidth]{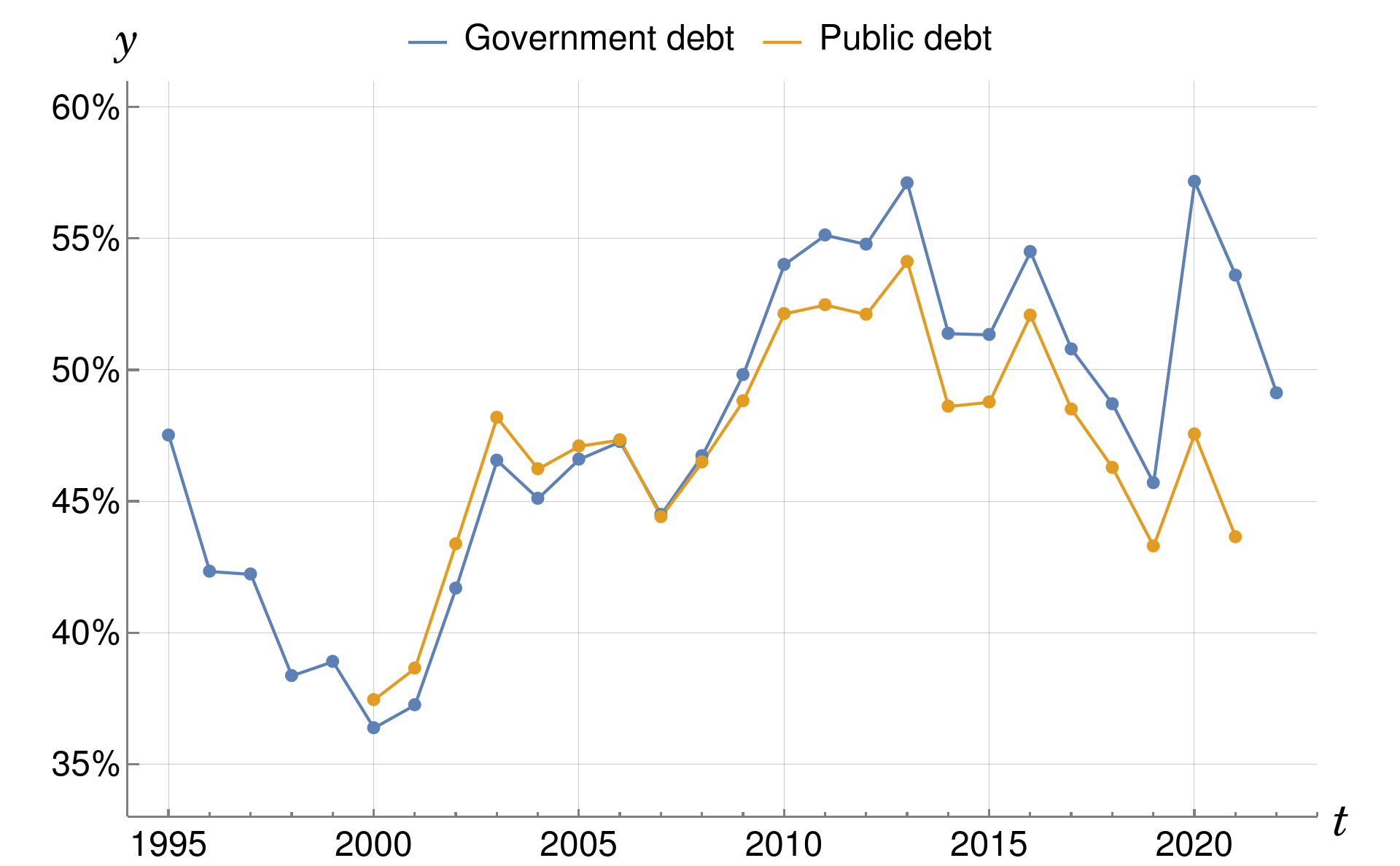}
    \caption{General government debt and total public sector debt relative to GDP.}
    \label{fig:debts}
\medskip
    \includegraphics[width=.925\textwidth]{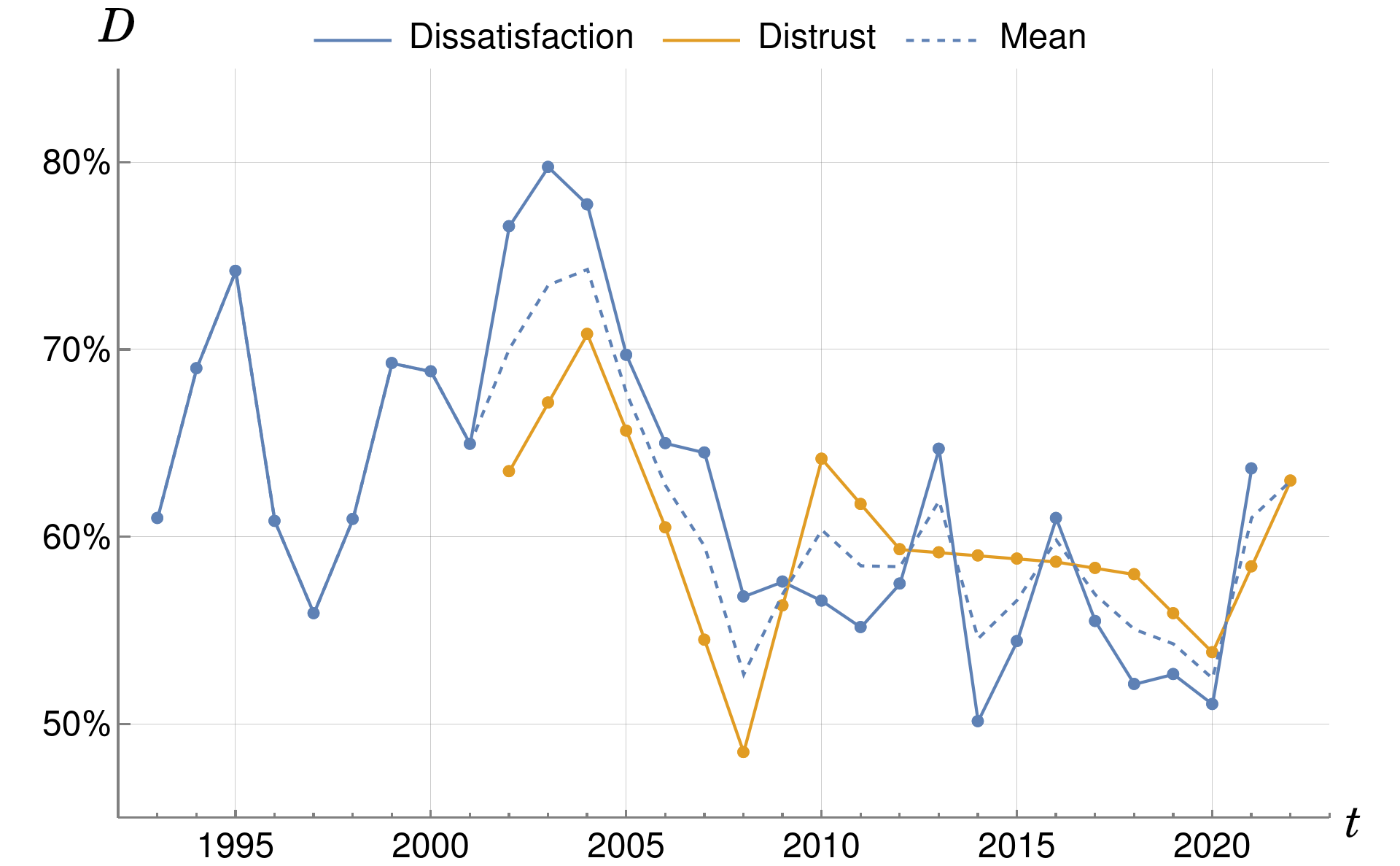}
    \caption{Two measures of distrust: dissatisfaction with democracy (blue), distrust in institutions (orange). The dashed line is their average (where it applies).}
    \label{fig:distrust}
\end{figure}

This quantity contains two quite different factors (whether they should be bundled will be discussed in later sections), and this is reflected in the data. The national debt is the least problematic. Although there are several quantities reported by GUS that could fit (public sector, institutions, local government etc.), their dynamics are tightly correlated, as seen in Figure \ref{fig:debts} so we used the general government debt (which includes local units), whose data go back to 1989. 

The distrust factor, on the other hand, presents the obvious difficulty of sparse or no data prior to 1991. The government-published opinion polls under the communist regime obviously cannot be trusted (or do not exist). And although there are foreign estimates of unrest like Cross-National Time-Series Data \citep{CNTS} they are either secondary (strikes, riots etc.), or in gross disagreement with the GUS data for post-communist years, e.g. the data on number of strikes \citep{Kramer}. Additionally, their widely varying criteria and components undermine a straightforward synthesis. In the end, we feel the most important flaw of using such measures is the follow: Since it is the role of $\Psi$ to predict the overall stress or unrest level, it would lead to a vicious circle if anything other than direct public opinion were to be used here -- we would be predicting unrest by measuring unrest.

We thus have a limited choice of: OBOP (later Kantar) polls, CBOS polls, GUS polls, and later European and other polls. Unfortunately, each spans different years, different (often changing) questions. For consistency's sake we will rely on the CBOS data for (dis)satisfaction with democracy and the their trust polls. Although Turchin originally phrased it as "Distrust in the government institutions", we are forced to piece a picture together from specific rating questions.

We propose to take the average of the following:
courts, government, parliament, police, public officials, and political parties -- they will constitute the distrust measure. The other variable is directly taken from the democracy satisfaction poll (Appendix~\ref{ListApp}).
In the period where they overlap, Figure \ref{fig:distrust}, they loosely follow the same trend; we will take as $D$ their mean for the years where they overlap, and the only one that was measured for the remaining periods.

\begin{figure}[!ht]
    \centering
    \includegraphics[width=.85\textwidth]{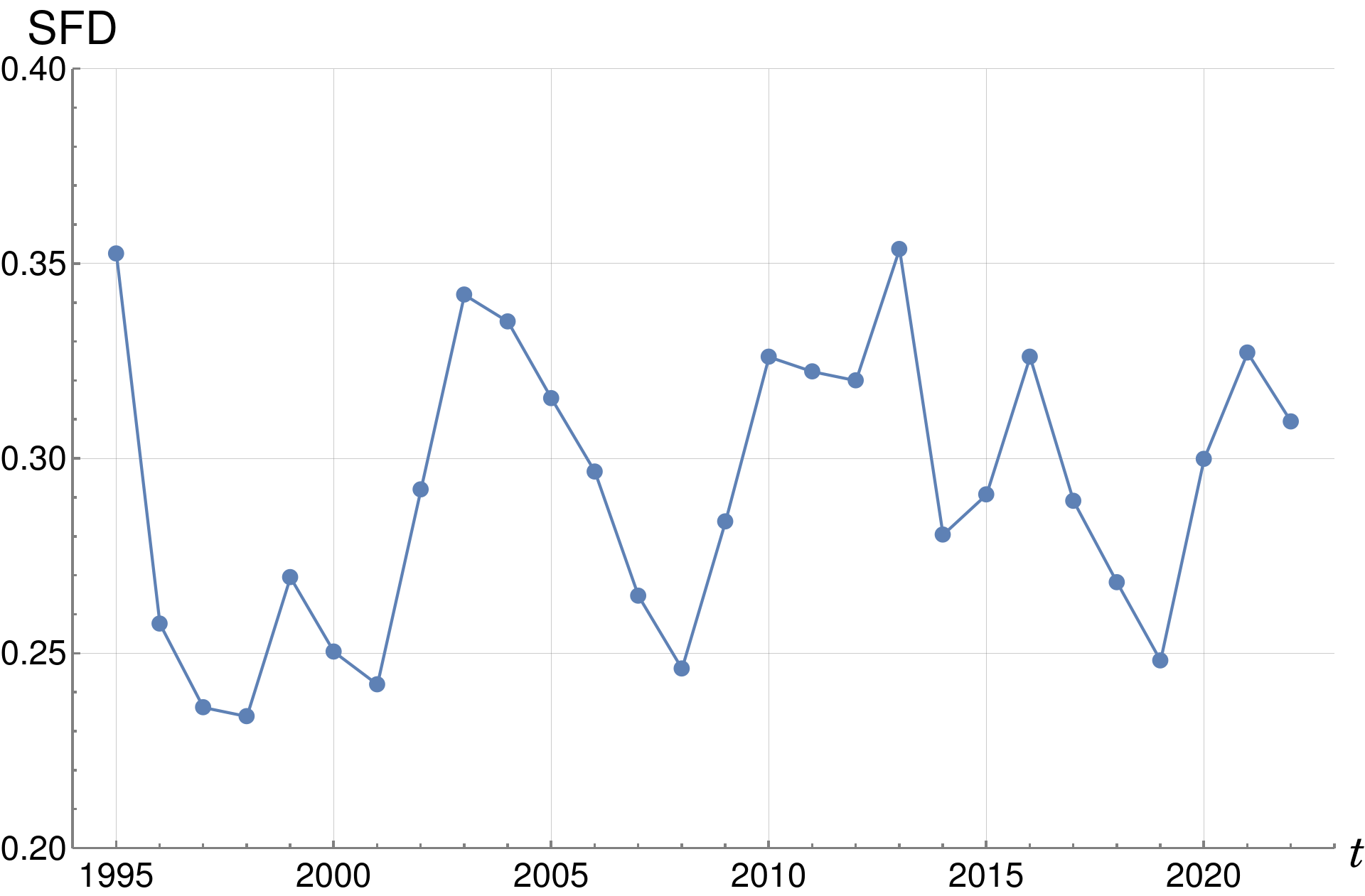}
    \caption{State Fiscal Distress.}
    \label{fig:SFD}
\end{figure}

Finally, the resulting SFD component is depicted in Figure \ref{fig:SFD}. Out of the three it appears to change the least, between 0.23--0.35, giving the range relative to its average as 0.41. For MMP and EMP the relative ranges are: 0.51 and 0.6, respectively.

Before jumping to the final discussion of PSI itself, we first take a historical detour, to try and reconstruct the indicators for a longer period, which lead to the emergence of Poland as a truly democratic country in 1989.

\section{Preliminary results: the recent years}

\begin{quote}
    \textit{,,I've always been careful never to predict anything that had not already happened.''}
\hspace*{\fill} Marshall McLuhan
\end{quote}

\begin{figure}[!ht]
    \centering
    \includegraphics[width=.85\textwidth]{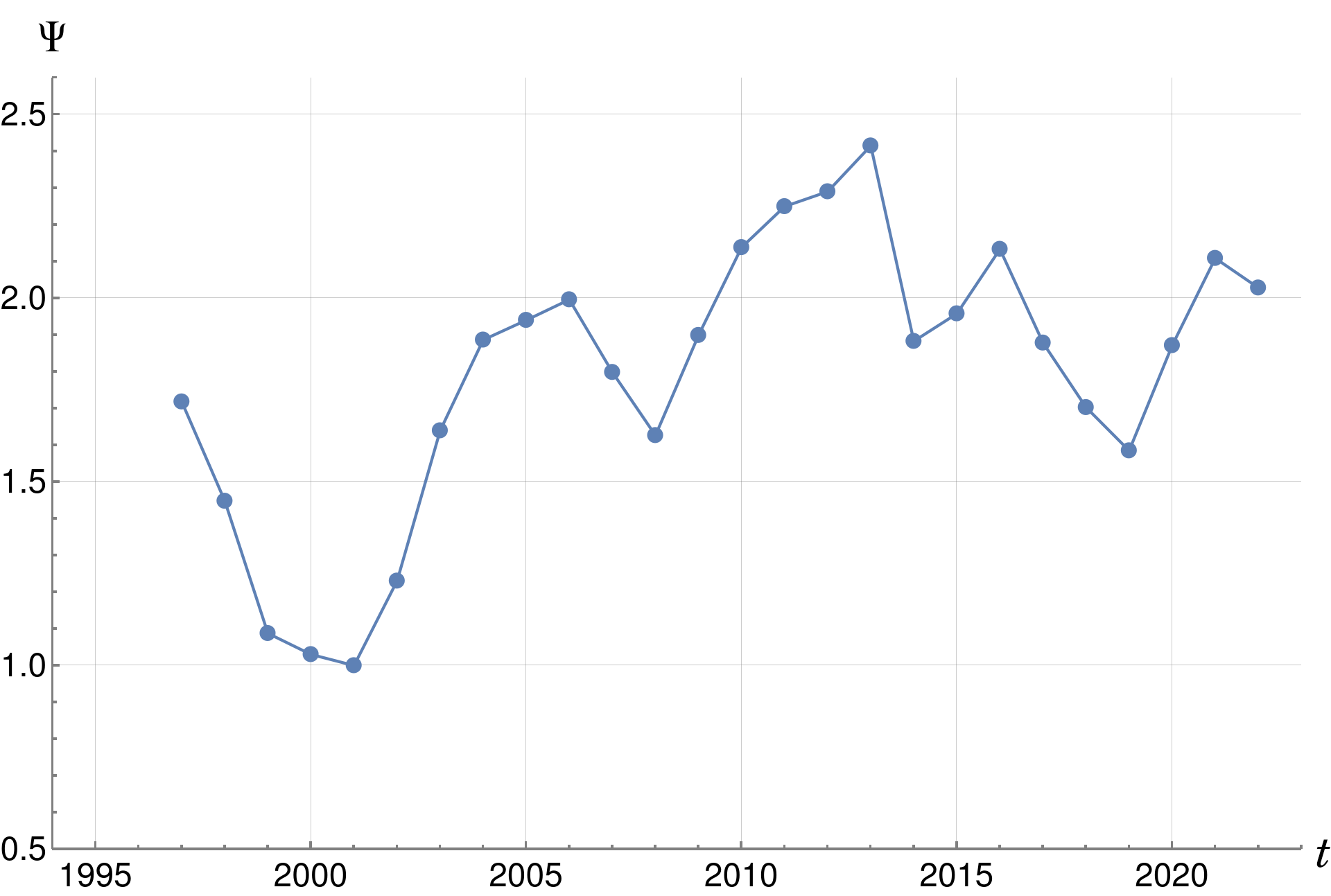}
    \caption{Political Stress Index obtained directly from data.}
    \label{fig:psi1}
\end{figure}
\begin{figure}[!ht]
    \centering
    \includegraphics[width=.85\textwidth]{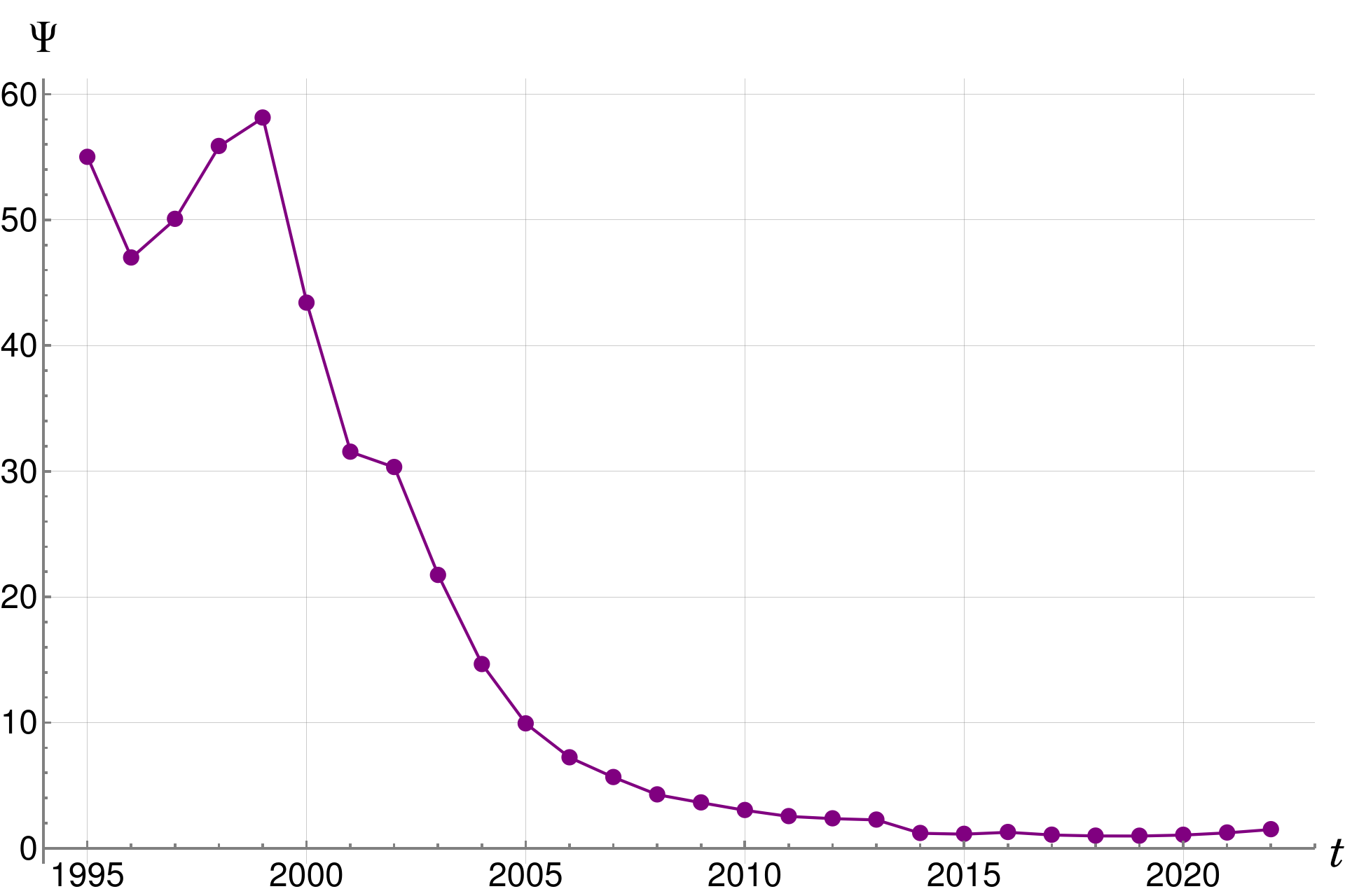}
    \caption{Political Stress Index, as obtained from the model.}
    \label{fig:psi2}
\end{figure}

We are now ready to present those results, that can be obtained from the most reliable, recent data in two versions: with the original elite model, and with data-based elite numbers. It must be said right at the outset, that \textbf{there is no indication of a revolt in the near future} in either version, though they show very different dynamics of PSI. 
This is in contrast to Turchin's connection between the American Civil War and the early XXI century \cite[graph 4.2]{Turchin:2013}, although we will see in the next section, that the model correctly identified the fall of communism in Poland.

It appears that both the data-based PSI (Figure~\ref{fig:psi1}) and the one obtained from the original model(Figure~\ref{fig:psi2}), could possibly be taken as post-revolutionary. The former is oscillatory, while the latter ``exponentially'' decreasing. We should be careful about this decrease, because it does not quite make sense when extended to the year 1950. That is the subject of the next section, and we will focus here on interpreting the data-based model in connection with the recent political history.

The fall of PSI after 1990 portrays the high hopes that the residents of the Eastern Block associated with the new-found freedom and transformations, which, nevertheless, did not benefit all classes equally due to the initially hasty privatisation and predatory capitalism.

Starting with 2010, PSI declines slightly. We consider this another indication of problems with the model's generalisability, at least to eastern Europe, as the index has failed to capture the ongoing increase in polarisation between the national right and liberal left fractions of the society. The recent public manifestations such as the Women's Marches, although not at the riot level, seem to contradict the PSI's stabilisation.

One could also make a case for interpreting the PSI around 2015 as depicting the transfer of power to PiS, a party whose social program can be classified as socialist: 500+ (fixed support per each child born), increasing both the level of retirement pensions, and their number ("thirteenth" month pension) etc., and whose campaign slogans about new, better redistribution of goods lead to a certain toning down of economic social tensions. At the same time, though, we have seen the rise of different kinds of tension -- symbolic and axiological. As in the American society, Poles remain deeply divided about ``pro life'' and ``pro choice'' -- the liberal left stands in opposition to the national, christian conservatives.

Finally, Poland entered the EU in 2004, and Schengen in 2007, which lead to a considerable increase in emigration. The youth bulge of 2006 was thus channeled outside, reducing the fuel necessary for outbursts of social unrest, riots or rebellion.

\section{Excursion into the Past}
\label{PRL}

\begin{quote}
\textit{Consolation by confabulation is the simplest stabiliser of social structures, whose incidental upside is that many dreadful expectations, based on the knowledge of depressing facts, do not come to pass after all, so by keeping those facts under the bushel, people are spared the stress.}\\
\hspace*{\fill} Stanis\l{}aw Lem, \textit{Wizja lokalna}
\end{quote}

\noindent As already indicated, the population data, and some of its structure, are readily available for much longer historical periods. Specifically: population ($N$), youth bulges ($A_{20-29}$), urbanisation ($N_{\text{urb}}$). Likewise, the wages are available going back to 1950.

\begin{figure}[ht]
    \centering
    \includegraphics[width=0.85\textwidth]{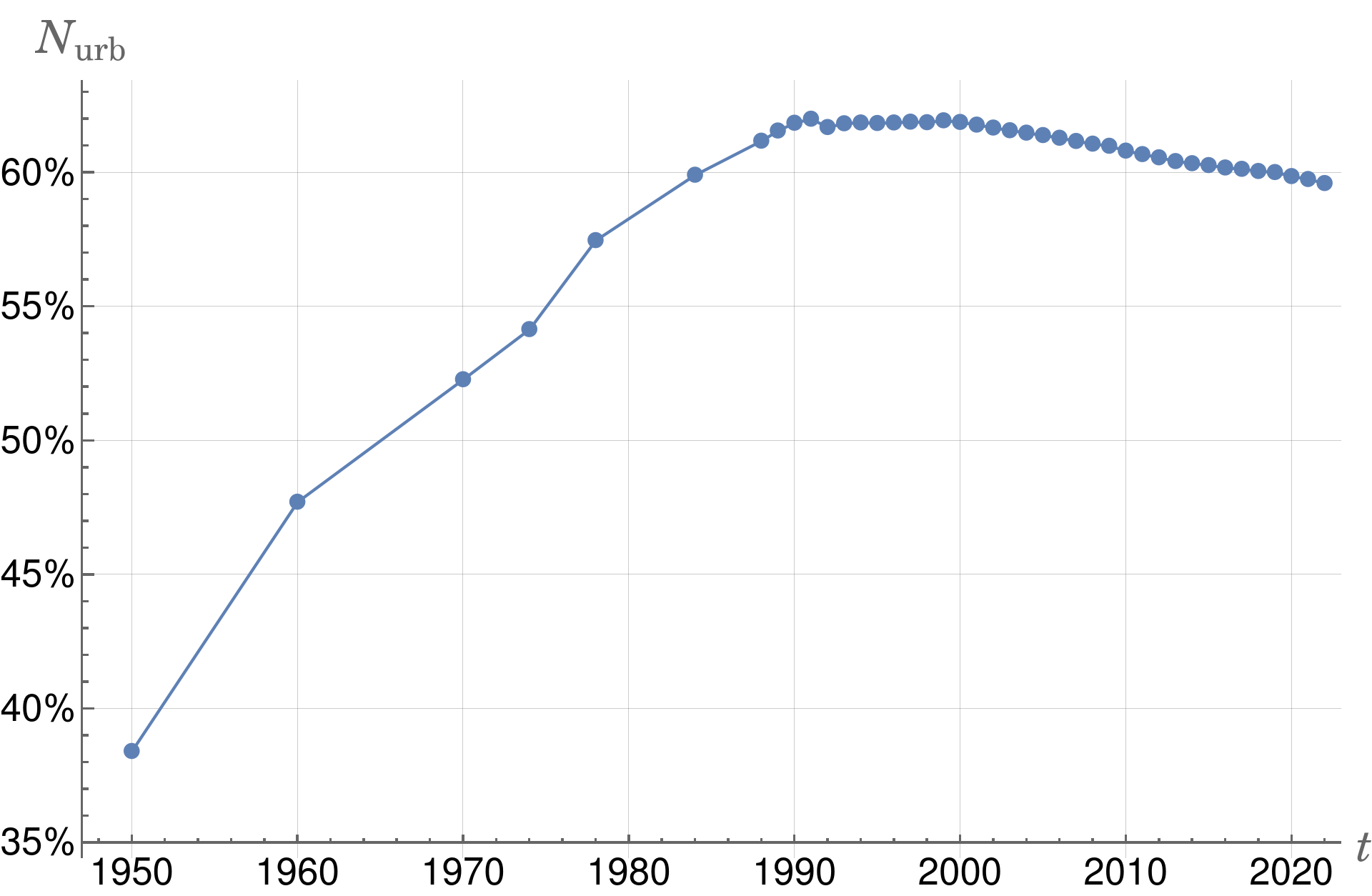}
    \caption{The percentage of the population living in cities (urbanisation) in the whole period analysed.}
    \label{fig:full_urb}
\end{figure}

Problems begin with elites -- their wages and classification of work groups. Although there exist archival annals of GUS, the categories and granularity of data vary too wildly to be used. Which is why we decided to take the basic mathematical model of section \ref{main_sec}, and use the derived $e$ and $\epsilon$.

\begin{figure}[h!t]
    \centering
    \includegraphics[width=0.95\textwidth]{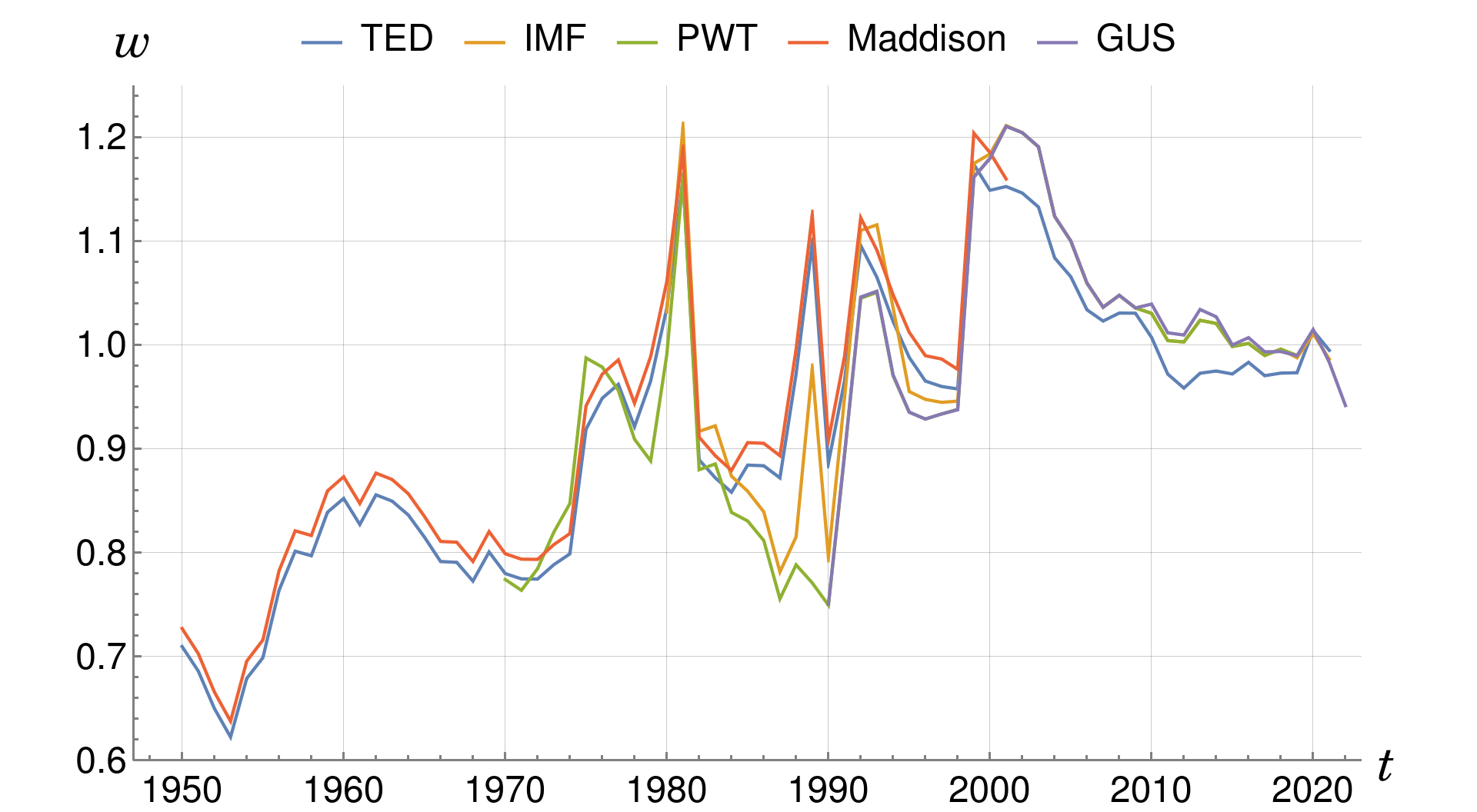}
    \caption{Historical wages relative to GDP per capita. GDP reconstructed using: Total Economy Database (TED, blue), International Monetary Fund (IMF, orange), Penn World Table (PWT, green), The World Economy \cite{Maddison} (Maddison, red), Central Statistical Office of Poland (GUS, purple).}
    \label{fig:full_wages}
\end{figure}

Things are not quite as hopeless when it comes to GDP and debt, as there are several sources, albeit with conflicting data. This is best seen in Figure \ref{fig:full_wages}, which shows several reconstructions of wages relative to the GDP per capita. There is only one version of the wages data, so the various $w$ differ only through the GDP. The index had to be reconstructed, as none of the sources give nominal values in PLN for the full period. Most often international dollars are used, recalculated for each specific year, with or without Purchasing Power Parity. To make them comparable with wages data, PPP had to be converted to current USD, then to current PLN, and finally, using the inflation data provided by GUS, to nominal prices. See Appendix \ref{ListApp} for sources of data, and conversion factors.

We chose to show $w$ rather than GDP itself simply because they both grow exponentially, and (even on a logarithmic scale) the differences are not as clearly visible for the raw variables as for their ratio. The differences lie only in the GDP data, as we consistently use the same wages data available since 1950.

The general refrain can be repeated again: since PSI is informative only relatively, and only when it changes by orders of magnitude, all the GDP reconstructions tell the same story. So, as with $D$ previously, we decided to take the mean of all measures available for any given year. 

\begin{figure}[h!t]
    \centering
    \includegraphics[width=0.85\textwidth]{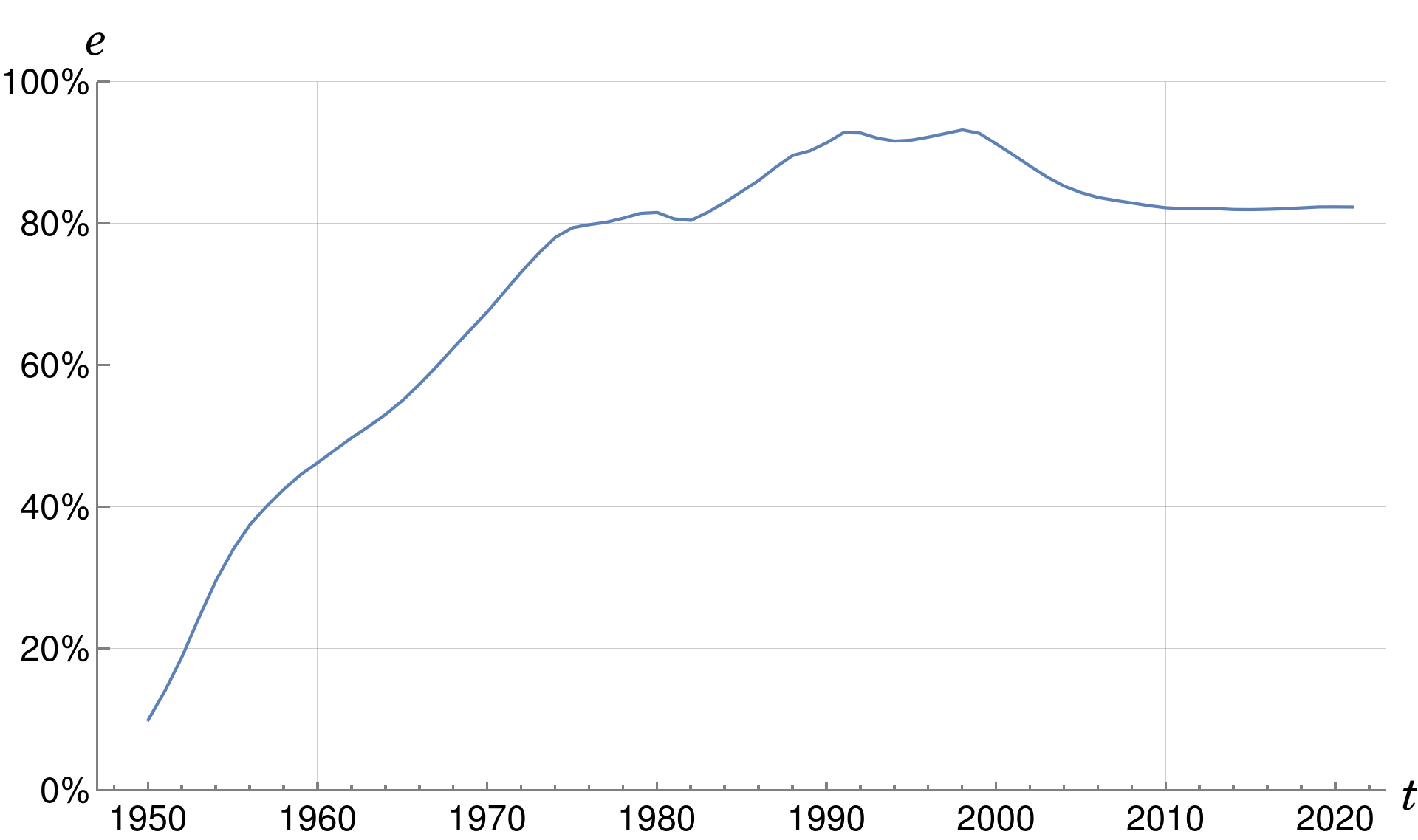}
    \caption{The percentage of elites in the population, as obtained from the model based on wages.}
    \label{fig:full_elites}
\end{figure}

Given the GDP and $w$, all other relative quantities and their derivatives follow; we show only the elite fraction $e$, Figure~\ref{fig:full_elites}, as it is the central point of contention.

Likewise, the national debt has been estimated by various agencies, methods and scholars. It exhibits the additional complication of the transfer rouble -- an artificial currency used in obligations between countries of the communist block. The exchange rates of roubles and of Polish zlotys were of course fixed centrally, and diverged from the more free and in some sense more realistic black market \citep{Kochanowski}. The wild behaviour is exemplified by the Penn World Table data, which gives exchange rates in the transition period as: 0.01132 in 1984, 0.1439 in 1989, and 0.95 in 1990 -- almost 2 orders of magnitude. This is further compounded with the inflation, which in 1990 reached the record level of 685.8\%.

Keeping this in mind, we have pieced together the national debt as follows: for the years 1950--1970 we followed \cite{Szpringer}, for 1970--1980 we used \cite{Kolodko}, for 1980--1990 \cite{Jachowicz} and IMF data, for 1990--2000 we took the reports of the Polish Ministry of Finance, and past 2000 the already mentioned GUS data. Wherever the data overlapped we took the average.

Finally, there is no reliable and continuous indicator or poll to measure public (dis)trust in the government, institutions or the state. Rather than create a patchwork approximation, we decided to exclude the variable $D$ altogether. This is not an optimal solution, but this way the scores between the past and the present become comparable. 

\begin{figure}[ht]
    \centering
    \includegraphics[width=.9\textwidth]{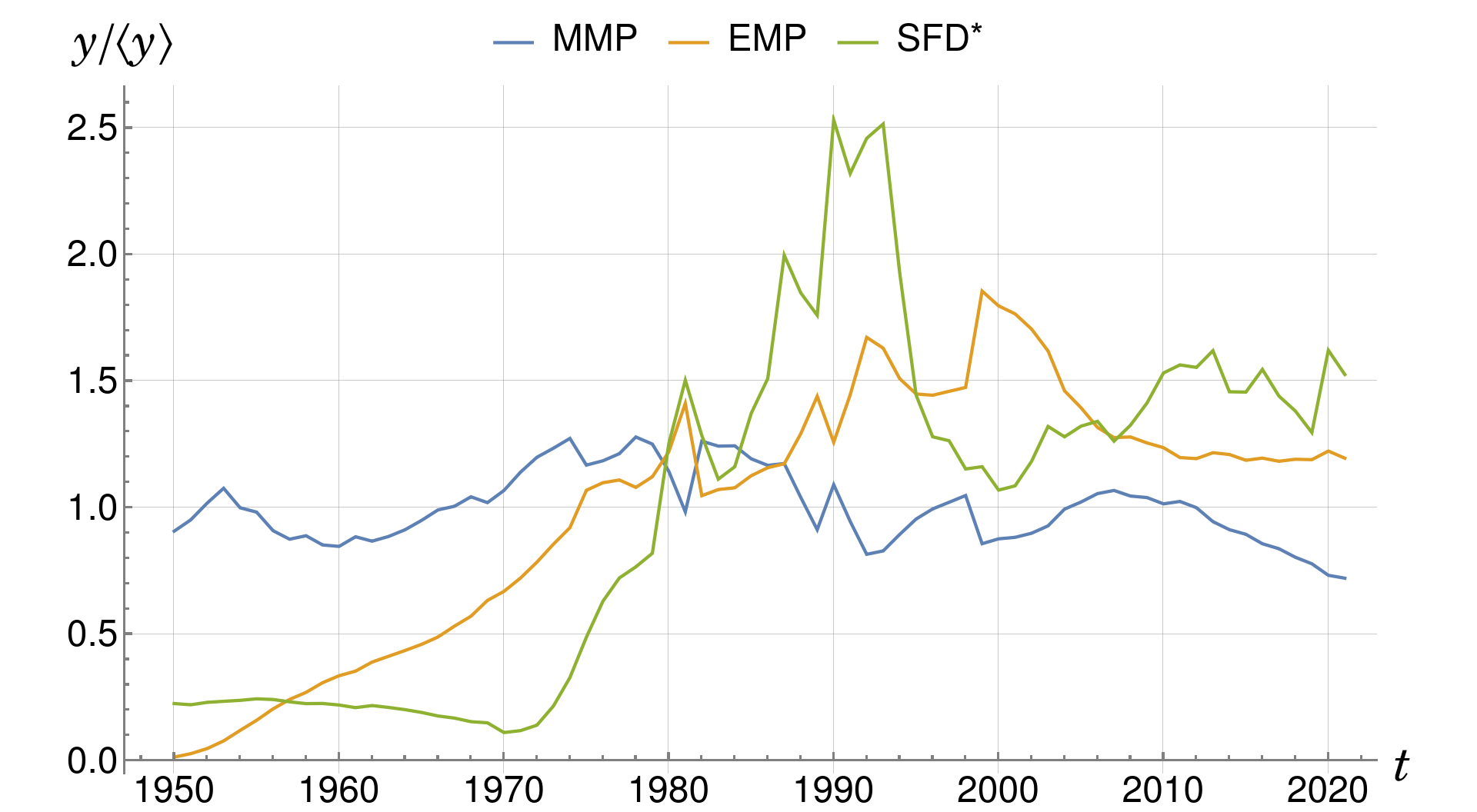}
    \caption{The three main components of PSI, rescaled by their mean (denoted as $y/\langle y\rangle$) for the whole historical period. SFD${}^*$ contains just the national debt.}
    \label{fig:PRL_comp}
\end{figure}

\begin{figure}[!ht]
    \centering
    \includegraphics[width=.85\textwidth]{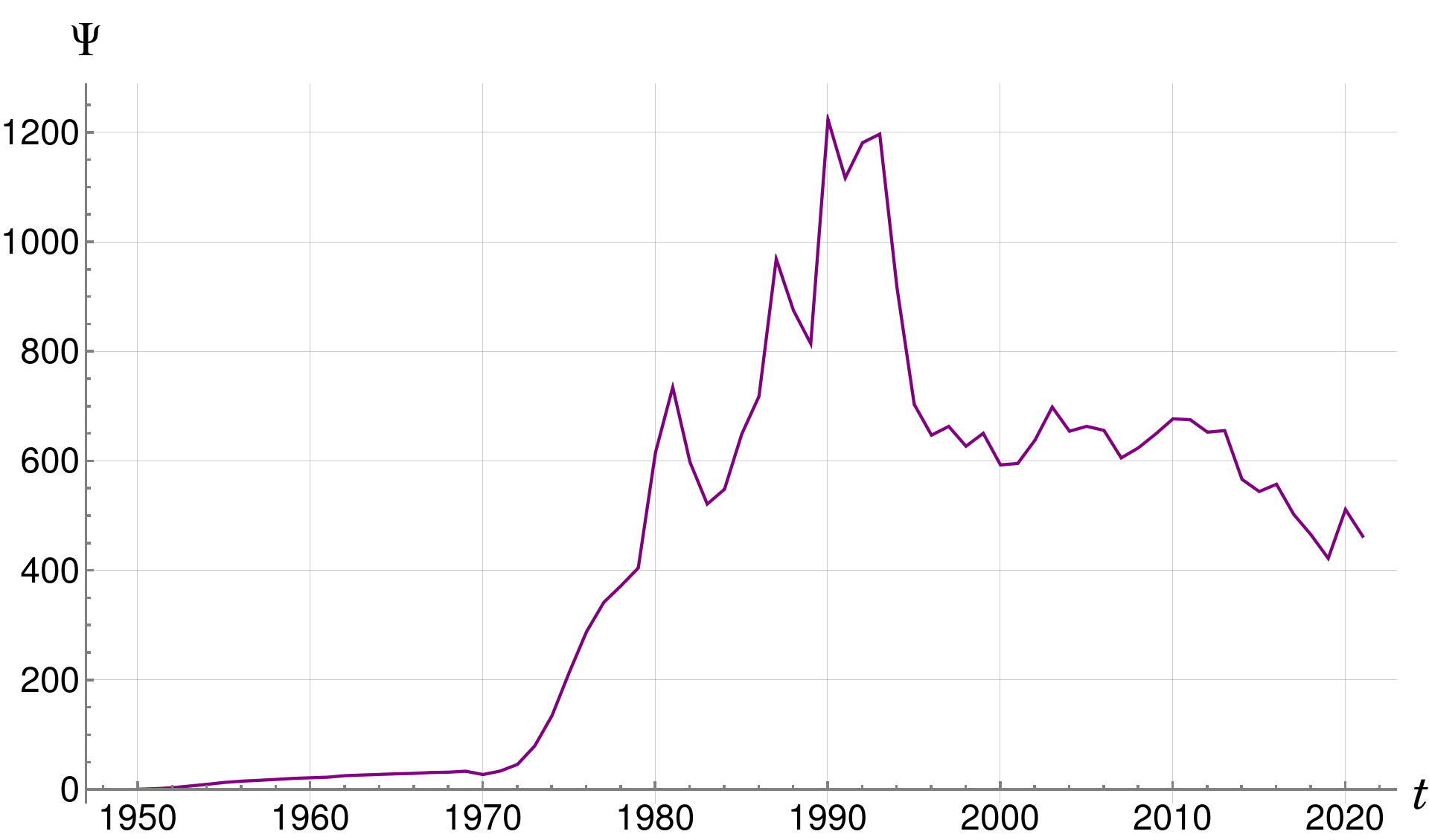}
    \caption{The Political Stress Index, for the whole analysed period. Elite components were obtained from the dynamical model.}
    \label{fig:PSI_PRL}
\end{figure}

This means that the relative debt $Y$ plays the role of SFD, and we plot all three main indicators (MMP, EMP, Y) together in Figure \ref{fig:PRL_comp}.
It is immediately striking, that the MMP is almost constant, while the debt increases 10 times, and EMP -- almost 200-fold. Those two components clearly dictate the final shape of PSI itself, shown in Figure \ref{fig:PSI_PRL}.


The main impression left by this final graph is the exceptional coincidence of the increase, and the maximum of $\Psi$ with the disintegration of the late communist regime. One should keep in mind though, that there is a considerable increase since 1950 to 1970 as well -- it just bleaks in comparison with what comes next; the data quality is also the worst in the earliest period.

Although the clear explosion of the index between the years 1970 and 1990 corresponds very well to history -- public debt in the 70's, the martial law of 1981, the Solidarity and the Round Table in 1989 -- it is unacceptable to think that the masses had negligible impact compared to intra-elite competition. So is the conclusion, that PSI remains 600 times higher now than in 1950 -- this is an artifact of the inadequate treatment of the elite fraction in the dynamical equations. 

Another undeniable feature of PSI's dynamics is that it very closely mimics that of the national debt: compare Figures \ref{fig:PRL_comp} and \ref{fig:PSI_PRL}. It adds up to the terrible fiscal state (high $y$) and poor living standards (low $w$) driving the discontent. But upward mobility was not an easy next step: in the state-owned industry of that time, the workers could not find better employment, as they would be paid the same everywhere. Nor could they privatise, and create the new upper-middle class. It is thus hard to interpret the growth of EMP (through $e$ and $\epsilon$) as swelling of the elites. When looking for the missing surplus GDP, we can instead point to the exploitative relation with the USSR \citep{Crane, Kalinski}, or to the inefficient administration, who acted according to Marxist ideals rather than economic theories -- even if money was spent on free, universal healthcare.

We can also now properly compare the present data and models with the past, but in stark contrast to the US, there is nothing in Fugers \ref{fig:psi1} and \ref{fig:psi2} to mimic the consistent growth in the 70's and 80's. We interpret this optimistically -- the model seems to indicate that no violent unrest or civil war is coming. At first glance this goes against the gloomy picture of social polarisation, protests and hate speech that the media served in the election year 2023, but the model does not include such variables by construction. We would risk the hypothesis, that those dimensions contributed to the transition in power in October 2023, well within the limits of the law, so PSI stayed low.

\section{A Closer Look at the Elite Growth Model}
\label{critique}
\begin{quote}
\textit{,,As scientists, our goal is not to save face, but in
fact to remove as much doubt as possible. Formal models make their assumptions
explicit, and in doing so, we risk exposing our foolishness to the world. This
appears to be the price of seeking knowledge. Models are stupid, but perhaps
they can help us to become smarter. We need more of them.''}
\hspace*{\fill}\citep{Stupid}
\end{quote}

The three main issues with the elite-mass interaction as described by equation \eqref{eq:ee}, are of rather different kinds: the first is mathematical in that the form of the equation introduces inconsistencies; the second adds the practical problem of finding realistic parameters, on which the model depends rather sensitively; finally the third is sociological -- it assumes specific reactions and motivations of people with regard to unsatisfactory wages. We shall consider them in more detail, proposing improvements whenever possible.

\subsection{The Dynamical Equations}
\label{sect:crit1}

Although equation \eqref{eq:ee} is supposed to govern the fraction of elites $e$, it is immediately visible that there are no natural bounds implied by the equation itself. In other words, a solution $e(t)$ might very well become greater than 1 or even worse -- negative. This does not require any especially nasty behaviour of the wages -- just a prolonged period of the (relative) wages staying below or above the reference level of $w_0$, respectively. To illustrate, the blue curves in Figure \ref{fig:params} are two such solutions (corresponding to real data).

\begin{figure}[ht]
    \centering
    \includegraphics[width=.85\textwidth]{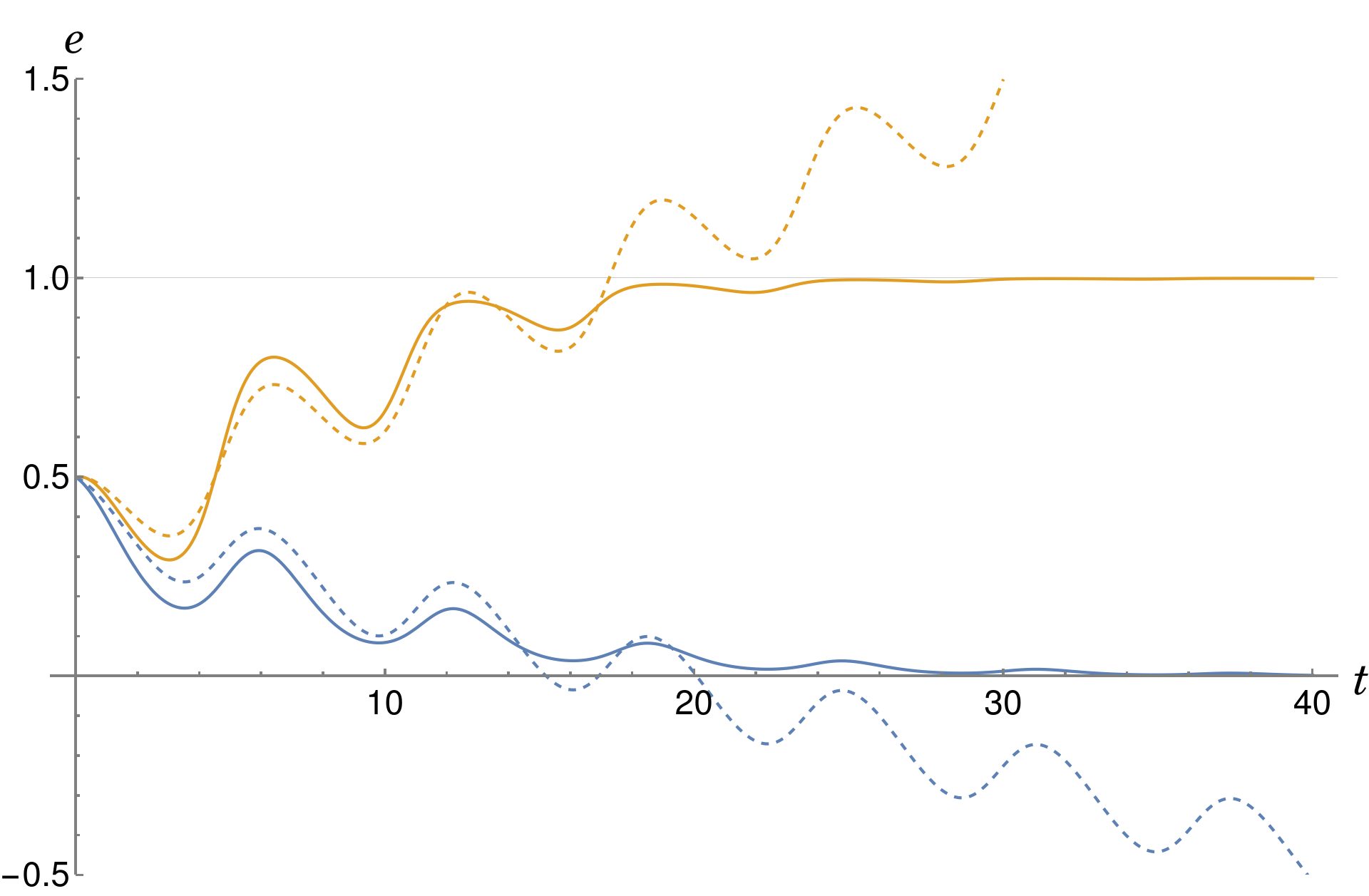}
    \caption{Evolution of the elite fraction with the logistic modification, $w_0=\mu_0=1$. Blue curve: $w = 1.12 + 0.33\sin(t)$; orange curve: $w=0.96+0.33\sin(t)$. The dashed graphs show the corresponding behaviour without the logistic factor.}
    \label{fig:logistic_mod2}
\end{figure}

One might expect that real wages will oscillate around $w_0$, making the positive and negative contributions cancel out. Unfortunately even that is not enough. The problem stems from the inversion of wages on the right hand side of the equation, solved in general as:
\begin{equation}
    e(t) = e(t_0) + \mu_0 \int_{t_0}^t \frac{w_0 - w(s)}{w(s)}\text{d}s.
    \label{eq:sol}
\end{equation}
If the integrand oscillates around zero, so will the integral, but that means the whole fraction, not just the wages $w$. The simplest counterexample is to take $w_0 = 1$ and $w(t) = 1 + a\sin(t)$ accordingly. The result, shown in Figure \ref{fig:logistic_mod2} demonstrates a catastrophic increase/decrease.

In general, if the relative wages can be approximated by a simple oscillation, there will be an exact connection between its amplitude and the reference wages level $w_0$ of the form:
\begin{equation}
    a^2 + w_0^2 = 1.
\end{equation}
at which $e(t)$ remains oscillatory. Any $w_0$ higher (lower) than the critical value
\begin{equation}
    w_* = \sqrt{1-a^2}
\end{equation}
will lead to the increase (resp. decrease) of $e$ in time. The detailed derivation of this fact can be found in Appendix \ref{CritApp}.

Thus the first obvious solution of this problem would be to adopt the critical level of $w_*$ as the reference at each point. But the difficulty is that in reality this will not be a constant, and only known for the past. It would also be doubtful that real wages oscillate with just one harmonic component $\sin(t)$.

A much surer, although still \textit{ad hoc}, modification is to force the differential equation itself to incorporate the natural boundaries. Just like in Lotka-Volterra systems, we are dealing with some sort of carrying capacity: elites cannot exceed 100\% (or fall below 0\%). It thus makes sense, to add a logistic factor:
\begin{equation}
    \dot{e} = e\, (1-e)\, \mu_0 \frac{w_0-w}{w}.
    \label{logistic}
\end{equation}

This is the most conservative form, as the carrying capacity is less than 100\% for the simple reason that if, say, 90\% of the society are in the group, they cannot be meaningfully called elite. Still, this completely solves the problem of runaway $e(t)$, regardless of whether $w_0$ has been set to the appropriate critical level. Figure \ref{fig:logistic_mod2} illustrates what happens, when there are exceeding fluctuations in $w$, with and without the logistic factor in equation \eqref{logistic}. For comparison, the original evolution is modelled with the logistic factor replaced by $1/6$, which is its average value.

Ultimately, more data and case studies are needed to shape the correct (or at least closer-to-truth) equation -- theorising will only get us so far. In the case of Poland, we did not need to introduce those corrections, because the data were mild enough to preclude negative values of $e$, with the right parameters -- but that is a problem in itself, discussed next. On the other hand, $e$ exceeding 100\% seems to be in line with the catastrophic increase of $\Psi$ as the harbinger of revolution, and we are inclined to think that it might be a feature rather than a bug of Turchin's model. Except that a society with even 90\% of elites sounds like Utopian science-fiction \citep{LeGuin}, and most certainly did not exist in communist Poland, as Figure~\ref{fig:full_elites} would suggest (even if those were unsatisfied elites). One solution is for $e$ to be reinterpreted, which we attempt in Section~\ref{sect:reint}. 

\subsection{The Dependence on Parameters}

One problem with the parameter $w_0$ was discussed above, and is structural. The other problem is the impact on the system's ``final'' state. No matter how small, the difference between the actual and critical values, decides between two diametrically opposing outcomes: the ``elite'' fraction tending to 0 or 100\% (or infinity in the naive model). In defense of the equations, it must be said, that this is nothing strange in a synthetic mathematical model, the insight here is simply, that $w_0$ is not a constant of Nature, but rather itself a dynamical variable -- people's desired standard of living changes due to attenuation.

Regarding $\mu_0$, we have already commented on the strange choice of values (0.1 and 0.002) in Section \ref{sect:EMP}, used without deeper justification in \cite{Turchin:2013}. Here we provide details on how much of an impact the parameter has on final results.

\begin{figure}[ht]
  \begin{subfigure}{0.495\textwidth}
    \centering
    \includegraphics[width=0.99\linewidth]{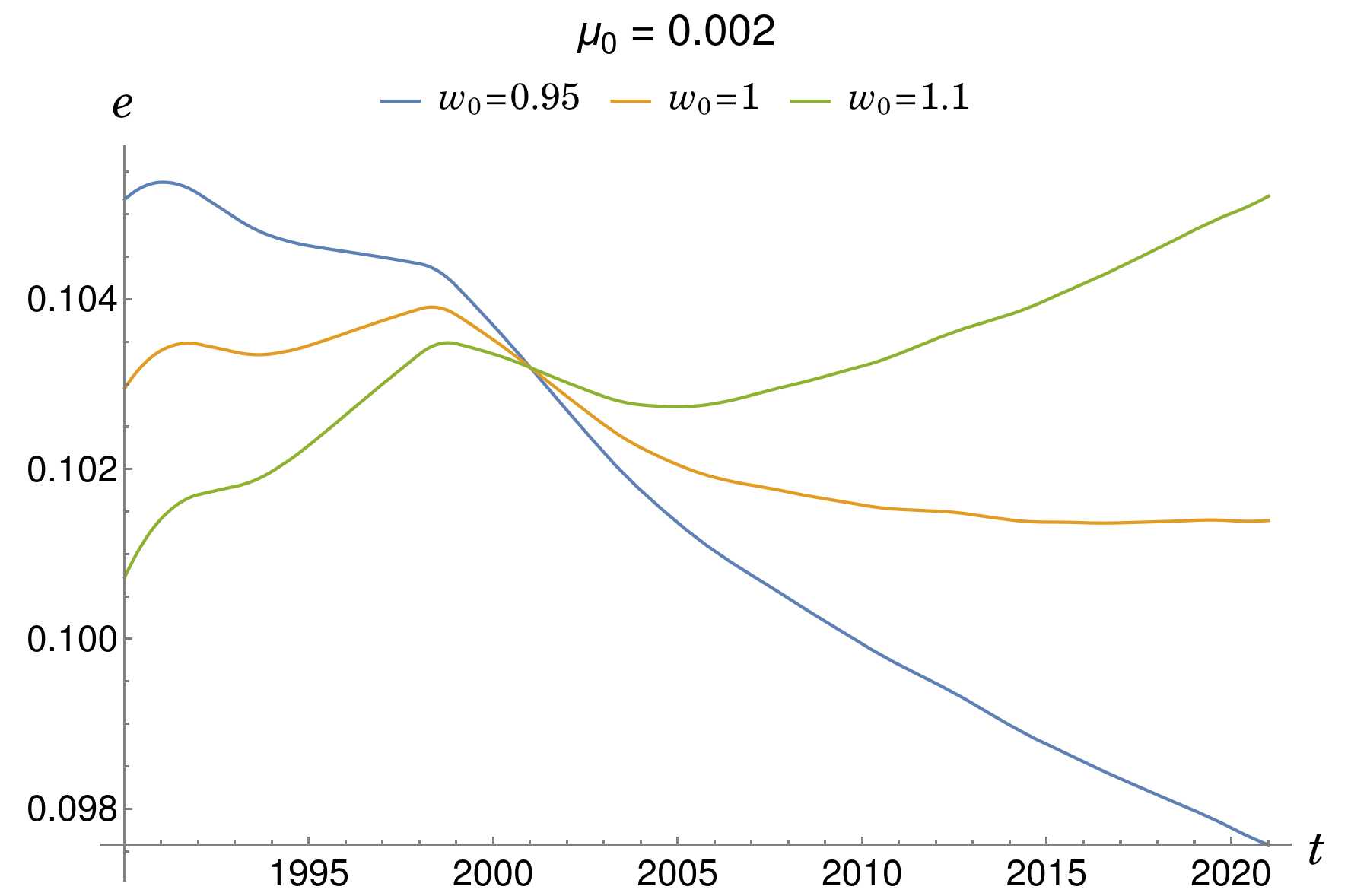}
  \end{subfigure}
  \begin{subfigure}{0.495\textwidth}
    \centering
    \includegraphics[width=0.99\linewidth]{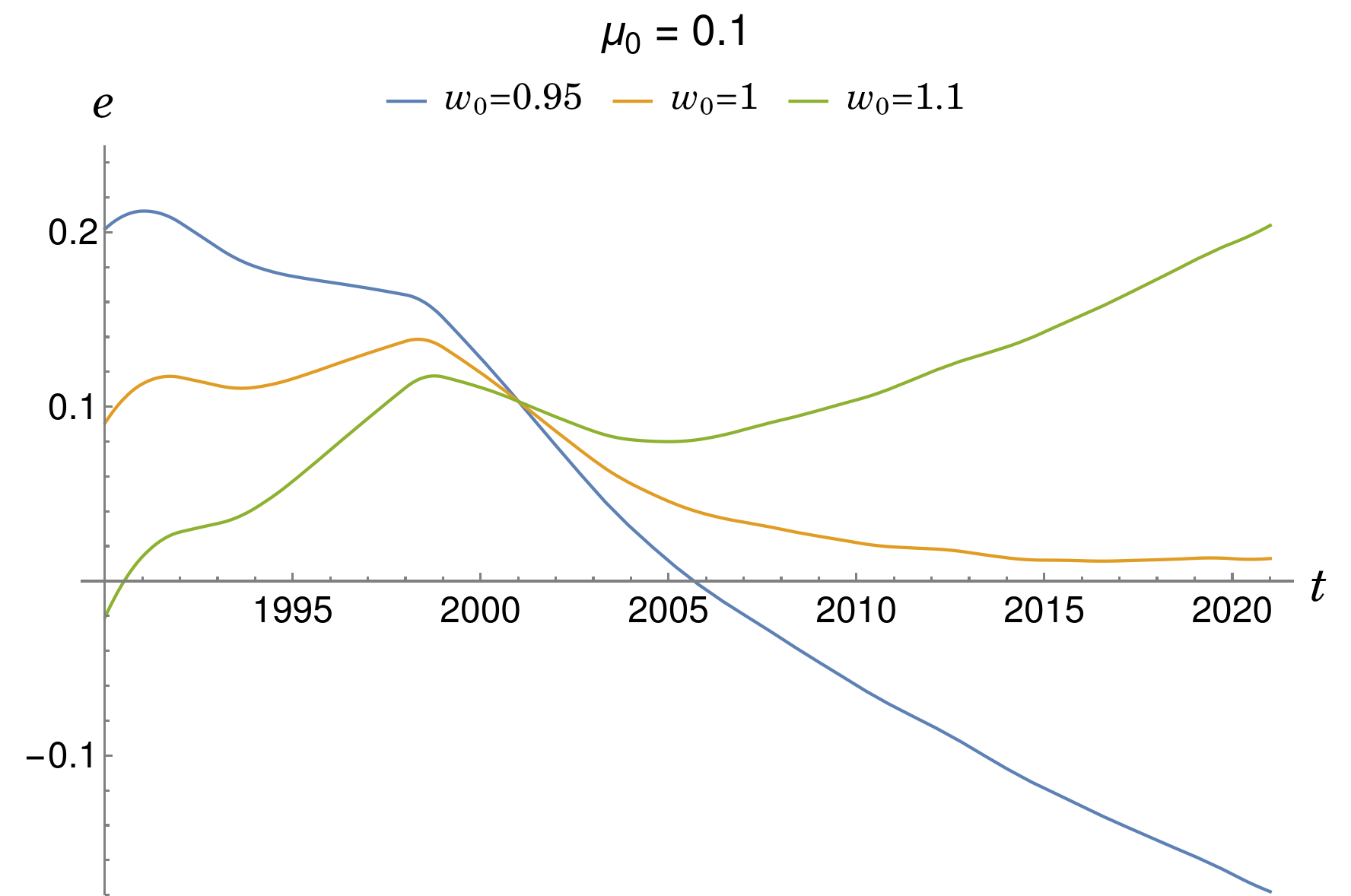}
  \end{subfigure}
    \caption{The elite fraction dynamics for different sets of parameters. The initial condition was chosen so that for all curves $e=10.32\%$ in the year 2001 (as obtained from ``standard'' model -- right, orange curve).}
    \label{fig:params}
\end{figure}

Although any scaling in $\mu_0$ only stretches the height of the graph linearly, this does not translate into linear growth in $e$. This happens because $\mu_0$ only multiplies the integral in \eqref{eq:sol}, whereas the starting point remains the same. This can be seen clearly in Figure \ref{fig:params}: the growth of the green curve between the years 1995 and 2020 is about $1.05/1.02 \approx 1.03$ on the left, whereas on the right it grows $0.2/0.05 = 4$ times. That is 3\% versus 300\% -- much too high to be caused simply by increasing $\mu_0$ 50 times -- and, more importantly, impossible to eliminate with any subsequent rescaling of $e$ itself. The reason is elementary: the effect of $\mu_0$ is \textit{multiplicative} on the drift accrued from the integral, whereas the initial value of $e$ is \textit{additive} when it comes to setting the reference level. Thus, the parameters need to be set carefully, as their interaction cannot be later removed with a simple division by the mean, or subtraction. What counts as ``reasonable'' would best be determined empirically, or at least with a deeper fundamental theory.

\subsection{The Work Ethic}
\label{work_ethic}

The last problem has to do with the values and goals of the masses in the context of perceived inequality, as expressed by the relative wages. It is by no means obvious, that the low wages will act as a motivation to try harder or retrain, and even less obvious that such a change is available ``on demand''. It is necessary both to appreciate the differences between countries, and include psychological and normative elements even when studying a single country.

The difference in work ethic between the US and Poland can be viewed in a wider cultural and historical context. The influence of the pervasive American ideal of WASP results in a distinct motivation of the masses, who, upon noticing their deteriorating material situation, redouble the effort to mend it. The protestant work ethic and the Calvinist predestination, induce a far stronger drive to exertion and improvement, than in catholic Poland, where the lack of worldly success generates attitudes of grievance, complaint and resignation instead of unleashing social and professional activism. This broad subject can also be traced back to Max Weber, but for modern results, we direct the reader to \cite{Arrunada} and references therein.

The pessimism of the Polish culture, regarding an individual's ability to overcome their plight has been decreasing only since the end of the 20th century, when capitalistic relationships started to influence the Sarmatian ethos of Polish intelligence \citep{Grzybowski, Cywinski}. Moreover, Poland's ``youth bulges'' (20--30 year-olds), react differently from those in the US, in situations perceived as unfavourable. They do not turn towards rebellion or revolution -- they emigrate. Young Americans have a much harder time finding a country to which they could escape to improve their overall situation.

\subsection{Reinterpretation of $e$}
\label{sect:reint}

No matter what mathematical solution is used to make the equations and variables consistent, the sociological problems of \ref{work_ethic} remain. The simulated elites do not reflect what we observe (in Poland), and neither is the drift mechanism convincing. And yet, the simple and satisfying trend of PSI (or even EMP alone) during the communist years in Poland suggests, that something real is being captured.

The majority of indicators making up PSI are financial, and so is $e$ by its very construction: it depends only on wage levels and GDP. Perhaps then its interpretation should depend on how it is constructed, rather than on a priori goals. The construction is special in that it integrates past years, and not just the momentary data -- loosely speaking, the indicator exhibits memory.

We would thus like to propose that what is really being accumulated in $e$ is the perceived discrepancy in wages, or more specifically, the \textit{resentment} of the masses due to over unsatisfactory earnings, accumulated over the years. If one needs to involve elites in it, it could also be thought of as envy of the elites, or rather of the perceived wages of the elites.

We note that the ``perception'' is important here, because comparison of wages in $e$ is not based on real elite income, as actual data shows. The division of surplus GDP among the elites is not the real mechanism, especially when it comes to elites that could be a conscious target for upward mobility: lawyers, doctors, or other high-level experts (like, most recently, programmers). No one can decide to change their job to ``millionaire'', an old-money aristocrat, let alone a backstage member of government. The job-switching decision must primarily be limited to jobs that are already included in macroeconomic statistics, and by extension in the usual relative wages $w$.

It follows, that the wage comparison would require subcategories of $w$, not manipulation of GDP. Unless, the comparison is not really between white- and blue-collar workers. If $e$ is reinterpreted as above, it makes sense to consider GDP per capita as a reference level, that is indirectly perceived by everyone, and with which everyone constantly compares their own life. If this comparison is unsatisfactory, the resentment/envy grows, but it takes many years to reach national levels of unhappiness. 

In particular, for the communist Poland, the discontent is very much fueled by the inability of an individual to enter real elites or even have any democratic impact on the government. Thus, without any mention of elite overproduction, the considerable growth of $e$ was still an obvious factor in the instability that would erupt in the 80's. Especially, as pointed out in Section~\ref{sect:crit1}, we would otherwise have to accept that 93\% of the late communist Poland population was in the elites (Figure~\ref{fig:full_elites}).

All in all, we consider the reinterpretation of $e$ to be the best immediate solution, as it would save the intuitive and promising features of the original model, while at the same time allaying some of the mathematical oversights.

\section{Conclusions}

As we tried to show there are too many independent problems with the implementation of the Goldstone model \citep{Goldstone} as proposed by \cite{Turchin:2013} to silently accept the apparent agreement with history. A fundamental revision and testing are needed to fully erase the impression of a just-so story.

This is mostly due to the mathematical shortcomings like inconsistent treatment of population fractions, or the unspecified parameter values and the model's sensitive dependence to them, which must definitely be addressed.
 Given the satisfactory agreement with Poland's historical transition, perhaps the main equations can be saved, and the suggested reinterpretation of the elites $e$ should be considered. 
 
Despite the \textit{post hoc} agreement with the fall of communism in Poland, the model of the elites fares poorly when compared with the actual elite data. This makes drawing conclusions for the country's future very risky. It does not invalidate the model as such, but 
at the very least, one has to account for
fine tuning and tacit assumptions only true in American conditions. With a view to generalisability, we can immediately point to:
\begin{itemize}
    \item The masses-elite interaction is very ``rigid'', i.e. it describes a closed system with one mode of interaction (mobility driven by wages), whereas for Polish citizens, as opposed to Americans, a wide array of emigration opportunities provides another such mode. This is especially significant in connection with the youth bulges, whose evolution has been greatly influenced by Poland's accession into the EU, and later the Shengen zone.
    \item The axiological dimension: racial tensions, or women's protests after the restrictions in abortion regulations, go beyond the financial concerns. In general, how people feel their values are being represented and protected goes beyond the basic political indicators of trust in the institutions, the state or democracy.
    \item Finally, the obvious point of considering a given country in isolation: the lack of accounting for external influences is especially questionable in view of the impact of the dissolution of the Soviet Union, Gorbachev's \emph{glasnost}. In this category lies also the very different geopolitical situation of each country: both the relation to its neighbours and the position in the respective continent.
\end{itemize}

Prominent experts in the theory of social conflicts, Lewis A. Coser \citep{Coser} and Ralf Dahrendorf \citep{Dahrendorf}, pointed out how hugely diverse the scale of social pressure (and intensity) is, and described the gradation of forms in which social unrest is expressed. Coser distinguished between ``real'' conflicts, focused on the problem of goods distribution, and ``unreal'' ones, which went beyond the material (they lasted longer, sometimes acting as substitutions for the former kind -- while still contributing to social stratification).

It would seem that the Goldstone/Turchin model does not address the latter type, as exemplified by the Women's Marches in Poland, after the restriction of abortion laws, or the wave of racial unrest initiated by the murder of George Floyd in the US in 2020, which undoubtedly increased social tension. Yet, as the comparison of Poland's past and present shows, these might be indicative of a different type of upheaval than a full-fledged civil war or uprising, for which PSI was tailored.

To briefly sketch some promising future research directions, we firstly wish to stress the need for extension of the model to include symbolic and axiological factors -- capturing the society's polarisation, which drives both intra- and inter-class conflict. This would serve to complement the ``culture'' factor of \cite{Turchin:2013}. Secondly, we feel that the role played by the social media cannot be forgotten -- both as a measure of said polarisation and tension, but also as a platform for conflict itself. Work on such measures has already been undertaken by others e.g. \cite{Azzimonti}, an example of an index of unrest as depicted in media can be found in \cite{Barrett:2020}. This might also offer a connection to treating social conflicts in a more structured way, i.e. describing their character, form and scale, going beyond simple (magnitude of) violence.

\section*{Appendices}
\appendix
\section{The Critical Parameter}

To derive the critical parameter for relative wages oscillating in the neighbourhood of $w=1$, let us assume the simplest cycle, i.e. with one harmonic (or Fourier) component:
\begin{equation}
    w = b + a \sin\left(\omega t \right),\quad 
    \omega = \frac{2\pi}{T},
\end{equation}
where $T$ is the period. This means, that the wage oscillate around $b$, with the amplitude of $a$, so that $w\in [b-a;b+a]$. The solution for the elite fraction $e$ is obtained by direct integration:
\begin{equation}
    e(t) = e(t_0) + \mu_0 \int_{t_0}^t \frac{w_0 - w(s)}{w(s)}\text{d}s.
\end{equation}
Because the integrand is periodic, the result will grow (or shrink) by the same amount each period, i.e
\begin{equation}
    \Delta e := e(T) - e(0) = \frac{\mu_0}{\omega} \int_0^{2\pi} 
        \left(\frac{w_0}{b+a\sin(\tau)}-1\right)\text{d}\tau,
\end{equation}
where the time variable was rescaled by $\omega$ for convenience.

The above difference must be zero for $e$ not to exhibit overall increase or decrease. The trigonometric integral can be calculated explicitly, giving the simple condition
\begin{equation}
    \Delta e = 0 \quad\Longleftrightarrow\quad
    2\pi\mu_0 \left(1-\frac{w_0}{\sqrt{b^2-a^2}}\right) = 0,
\end{equation}
Solving it, we obtain the sought for definition of the critical parameter $w_*$ to be
\begin{equation}
    w_*^2 = \sqrt{b^2-a^2}.
\end{equation}

\label{CritApp}
\section{Data Sources}
\label{ListApp}

\begin{itemize}[leftmargin=0mm]
\item {\bf Population:}\\
GUS, annual macroeconomic indicators\\
\url{https://stat.gov.pl/wskazniki-makroekonomiczne/}\\
GUS, emigration data\\
\url{https://stat.gov.pl/obszary-tematyczne/ludnosc/migracje-zagraniczne-ludnosci/glowne-kierunki-emigracji-i-imigracji-na-pobyt-staly-w-latach-1966-2023,4,2.html}\\
GUS, emigration by age group\\
\url{https://dbw.stat.gov.pl/baza-danych}

\item {\bf Gross Domestic Product:} GUS, Macroeconomic Databank\\
\url{https://bdm.stat.gov.pl/}

\item {\bf Historical GDP:}\\
    Total Economy Database \url{https://www.conference-board.org/data/economydatabase/total-economy-database-methodology}\\
    International Monetary Fund \url{https://www.imf.org/en/Publications/WEO/weo-database/2022/October}\\
    Penn World Table \url{https://www.rug.nl/ggdc/productivity/pwt/pwt-releases/pwt100}\\
    The World Economy by \cite{Maddison}

\item {\bf Wages:} GUS, as tabulated by ZUS.\\
\url{https://www.zus.pl/en/baza-wiedzy/skladki-wskazniki-odsetki/wskazniki/przecietne-wynagrodzenie-w-latach}

\item {\bf Elite numbers and wages:} The structure of average gross wages and salaries of paid employment by PKD/NACE sections, as found in GUS Yearbook of Labour Statistics 1998--2021.

\item {\bf Youth bulges: } Age pyramid of GUS, with the year 2021 corrected.
\item {\bf Unemployment: } GUS, annual macroeconomic indicators.\\
\url{https://stat.gov.pl/wskazniki-makroekonomiczne/}
\item {\bf Government debt:} GUS, annual macroeconomic indicators.\\
\url{https://stat.gov.pl/wskazniki-makroekonomiczne/}\\

\item{\bf Historical debt:}\\
World Economic Outlook Database, International Monetary Fund\\
\url{https://www.imf.org/en/Publications/WEO/weo-database/2022/October}\\
``Public foreign debt in Poland in historical perspective'' \cite{Szpringer} (in Polish)\\
``Crisis, Adaptation, Development'' \cite{Kolodko} (in Polish)\\
``Poland in Debt Crisis'' \cite{Jachowicz} (in Polish)\\
Public Debt -- yearly report 2001, Polish Ministry of Finance (in Polish)\\
\url{https://www.gov.pl/web/finanse/raport-roczny-archiwalne}

\item {\bf Distrust: } CBOS, series of polls: "Polacy o demokracji" (Poles about democracy) and "Zaufanie spo\l{}eczne" (public trust).\\
\url{https://www.cbos.pl}

\item{\bf Inflation:} GUS\\
    \url{https://stat.gov.pl/obszary-tematyczne/ceny-handel/wskazniki-cen/wskazniki-cen-towarow-i-uslug-konsumpcyjnych-pot-inflacja-/}

\end{itemize}

\end{document}